\documentclass[12pt]{article}
\usepackage{a4,amsmath,amssymb,amsthm,cite,graphicx,slashbox,cancel,hyperref}
\usepackage[table]{xcolor}
\usepackage{tikz}
\usepackage{ifthen,tikz-3dplot}
\usetikzlibrary{intersections,calc,matrix,arrows, decorations.markings,shapes}
\def\Bbb{\mathbb}
\def\BZ{\Bbb Z} 
\def\BC{\Bbb C} 
   \def\BH{\mathbb{H}}
 \def\BN{\mathbb{N}}

\def\Tr{\mathrm{Tr}}

\newcommand{\Sp}{\textrm{Sp}}
\newcommand{\etabox}[2]{\underset{\ ~#2}{#1\ \framebox[15pt]{\phantom{a}}}}
\catcode`@=11 \@addtoreset{equation}{section} \catcode`@=12

\newtheorem{theorem}{Theorem}[section]

\begin{document}
\bibliographystyle{utphys}
\begin{titlepage}
\renewcommand{\thefootnote}{\fnsymbol{footnote}}
\noindent
\mbox{}\hfill
{\tt arXiv:1905.06083 [hep-th]} \\[4pt]
\mbox{}\hfill 
\hfill{\fbox{\textbf{v2.0; Sept 2019 }}}

\begin{center}
\large{\sf  BKM Lie superalgebras from counting twisted CHL dyons -- II }
\end{center} 
\bigskip 
\begin{center}
Suresh Govindarajan\footnote{\texttt{suresh@physics.iitm.ac.in}}  and Sutapa Samanta\footnote{\texttt{sutapa@physics.iitm.ac.in}}\\
\textit{Department of Physics, Indian Institute of Technology Madras,\\ Chennai 600036, INDIA.}
\end{center}
\bigskip
\begin{abstract}
We revisit earlier  work by one of us which lead to a periodic table of Borcherds-Kac-Moody algebras that appeared in the context of the refined generating function of quarter-BPS states (dyons) in $\mathcal{N}=4$ supersymmetric four-dimensional string theory. We make new additions to the periodic table by making use of connections with generalized Mathieu moonshine as well as umbral moonshine. We show the modularity of some Siegel modular forms that appear in umbral moonshine associated with Niemeier lattices constructed from A-type root systems  and further show that the same Siegel modular forms appear for generalized Mathieu moonshine in some cases. We argue for the existence of a new kind of BKM Lie superalgebras that arise from the dyon generating functions for the $\BZ_5$ and $\BZ_6$ CHL orbifolds.
\end{abstract}
\end{titlepage}
\setcounter{footnote}{0}

\clearpage
\tableofcontents
\section{Introduction}

The counting of BPS states in $\mathcal{N}=4$ supersymmetric four-dimensional string theories has proved to be a playground that has lead to interesting and unexpected  connections between automorphic forms, finite simple groups as well as Lie algebras. In these theories, the generating function of half-BPS states is a modular form while the one for quarter-BPS states is a Siegel modular form\cite{Dijkgraaf:1996it,Sen:2005ch,Jatkar:2005bh}. An unexpected connection to the largest sporadic Mathieu group $M_{24}$ appeared in this counting problem\cite{Govindarajan:2009qt,Eguchi:2010ej}. This is referred to as Mathieu moonshine. Connections with Borcherds Kac-Moody (BKM) Lie superalgebras also emerged\cite{Cheng:2008fc,Govindarajan:2008vi,Cheng:2008kt}.

In \cite{Govindarajan:2010fu} (referred to as Paper 1 in the sequel), one of us constructed a `periodic table of BKM Lie superalgebras' by associating a BKM Lie algebra to a pair of commuting elements of $M_{24}$. All the examples that arose in this paper appeared independently in the construction of `dd-modular' forms by Cl\'ery and Gritsenko\cite{Gritsenko:2008}. This paper aims to extend the results of Paper 1 in several ways: (i) Extend the examples considered and (ii) make connections with developments such as umbral moonshine\cite{Cheng:2012tq}. \\

\noindent A summary of our results is as follows.
\begin{enumerate}
\item We provide new formulae for Siegel modular forms that arise as the additive lift for Jacobi forms of $\Gamma_0(q,Nq)$ -- these are the S-transform of the additive lift of Cl\'ery and Gritenko. This directly provides the sum side of the Weyl-Kac-Borcherds (WKB) denominator formula for Chaudhuri-Hockney-Lykken(CHL) $\BZ_N$ orbifolds\cite{Chaudhuri:1995fk}.
\item Using a Hecke operator given in \cite{Govindarajan:2011em} for generalized moonshine, we obtain product formulae for Siegel modular forms. This provides a uniform description for the product side of the WKB denominator formula for all examples. In some cases, these are equivalent to the Borcherds product formula of Cl\'ery and Gritsenko described in Theorem \ref{CGproduct}.
\item We provide the modularity of  Siegel modular forms that arise with umbral moonshine associated with  $A$-type Niemeier root systems. The zeroth Fourier-Jaocobi coefficient which is also the seed for the additive lift when it exists, is also determined.  We also find two new examples that correspond to twining by order two and three elements of $2.M_{12}$ at lambency three. Some of these Siegel modular forms also arise in the context of generalized Mathieu moonshine.
\item Contrary to the claim in \cite{Krishna:2010gc}, we show that the additive lift for the $\BZ_5$ orbifold explicitly agrees with the product formula.  This provides Macdonald type identities for the $\BZ_5$ and $\BZ_6$ CHL orbifolds.  We argue that these identities provide evidence for the existence of a new class of BKM Lie algebras associated with hyperbolic lattices with Weyl vector of hyperbolic type. 
\item We add two new rows and two columns to the periodic table of BKM Lie superalgebras given in Paper 1.
\end{enumerate}

The organisation of this paper is as follows. After the introduction section, section 2 reviews the connection between walls of marginal stability and rank three Lorentzian lattices. We also set up the notation to make connection with the data that is input in the construction of  Lorentzian Kac-Moody Lie superalgebras of Gritsenko and Nikulin. In section 3, we provide direct construction of various Siegel modular forms so that their modular properties are evident and their connection with generalized Mathieu moonshine is manifest. We also provide a uniform construction of Siegel modular forms associated with umbral moonshine associated with  Niemeier lattices constructed from A-type root lattices. This is summarised in Table \ref{umbraltable}. In section 4, we make the connection with BKM Lie superalgebras extending results of paper 1. Even for the cases that already appeared in paper 1, we provide a uniform description. We provide some evidence for the existence of new BKM Lie superalgebras that extend Lorentzian Kac-Moody Lie superalgebras of Gritsenko and Nikulin to include Lorentzian root lattices with Weyl vector of hyperbolic type. This is done with three examples associated with the Cartan matrices $A^{(5)}$ and $A^{(6)}$. The results are summarised in an updated periodic table of BKM Lie superalgebras (see Table \ref{periodictable}). We conclude in section 5 with some remarks. Appendix A deals with the details of the paramodular group and its subgroups. Appendix B provides three different constructions of Siegel modular forms that we use in this paper. In appendix C, we provide some details of a computation.

\section{Root Systems from walls of marginal stability}\label{SecLattice}

We shall focus on CHL models with $\mathcal{N}=4$ supersymmetry obtained as the $\mathbb{Z}_N$ orbifolds of the heterotic string compactified on $T^6$ for $N\in\{1,\ldots,6\}$.  We discuss the hyperbolic root systems of rank three that arise from the walls of marginal stability in these  orbifolds. 

\subsection{Walls of marginal stability and hyperbolic polygons}

The walls of marginal stability (where two-centred black holes decay) in these CHL models  are determined by its intercepts on the real part of the upper half plane (which is the moduli space of the heterotic dilaton-axion fields)\cite{Sen:2007vb}. The precise curve (wall) however depends on other moduli fields. These walls determine a hyperbolic polygon in the upper half-plane. It is anticipated that there is a Lie algebra with a  Weyl chamber whose intercepts are identical to those of the hyperbolic polygon\cite{Cheng:2008fc,Cheng:2008kt,Govindarajan:2009qt}. Let us denote this polygon by $\mathcal{M}^{(N)}$. The polygon for $N=1,2,3$ is given by set of vertices $\mathcal{V}_N$:
\begin{equation}
 \mathcal{V}_1 = \left(\tfrac11,i\infty, \tfrac01, \tfrac11 \right)\ ,\  \mathcal{V}_2 = \left(\tfrac12, \tfrac11,i\infty, \tfrac01, \tfrac12\right) \ ,\ 
 \mathcal{V}_3 = \left(\tfrac12, \tfrac23, \tfrac11, i\infty, \tfrac01, \tfrac13,\tfrac12\right) \  .
\end{equation}
When $N>3$, the polygons have infinite edges. It is preferable to use an algebraic method to describe the polygon. For $N\in\{1,2,\ldots,6\}$, let $\lambda_N$ denote any root of the quadratic equation \[\lambda^2 - (N-2) \lambda +1=0\ .\] 
Define
\[
\mathcal{A}^{(N)}_m = \begin{cases}
\frac{\lambda_N^m -\lambda_N^{-m}}{\lambda_N -\lambda_N^{-1}}\ , & \text{ for }N\neq 4 \\
m\ , & \text{ for } N=4 
\end{cases}
\quad,\quad
\mathcal{B}_m^{(N)}=\mathcal{A}^{(N)}_m + \mathcal{A}^{(N)}_{m+1}\ .
\]
Form the ordered sequence
\[
\mathcal{V}_N = \left(\frac{\mathcal{B}_{m-1}^{(N)}}{N\mathcal{A}^{(N)}_m} , \frac{\mathcal{A}^{(N)}_m}{\mathcal{B}_m^{(N)}}\quad, \text{ for } m\in \mathbb{Z}\right)\ .
\]
Explicitly one has
\begin{align}
\mathcal{V}_4 &= \left(\tfrac12,\ldots,\tfrac58,\tfrac23,\tfrac34,\tfrac11,i\infty,\tfrac01,\tfrac14, \tfrac13,\tfrac38,\ldots,\tfrac12 \right)\ , \nonumber \\
\mathcal{V}_5 &= \left(\tfrac12\left(1+\tfrac1{\sqrt5}\right),\ldots,\tfrac{11}{15},\tfrac34,\tfrac45,\tfrac11,i\infty,\tfrac01,\tfrac15, \tfrac14,\tfrac4{15},\ldots,\tfrac12\left(1-\tfrac1{\sqrt5}\right) \right) \ ,\\
\mathcal{V}_6 &= \left(\tfrac12\left(1+\tfrac1{\sqrt3}\right),\ldots,\tfrac{19}{24},\tfrac45,\tfrac56,\tfrac11,i\infty,\tfrac01,\tfrac16, \tfrac15,\tfrac5{24},\ldots,\tfrac12\left(1-\tfrac1{\sqrt3}\right) \right)\ . \nonumber
\end{align}
We indicate the limit points of the sequence at the start and end of the sequences. For $N=4$, since the two end-points are equal to $\frac12$, we obtain a closed polygon.
For $N=5,6$, the polygons are open. We need to add a second set of vertices and edges to get a closed polygon. 
\[
\widetilde{\mathcal{A}}^{(N)}_m = \frac{\lambda_N^{m} -\lambda_N^{-m}-\lambda_N^{m-1}+\lambda_N^{1-m}}{\lambda_N -\lambda_N^{-1}}\quad,\quad
\widetilde{\mathcal{B}}_m^{(N)}=\lambda_N^{m}+\lambda_N^{-m}\ ,
\]
Form the ordered sequence
\[
\widetilde{\mathcal{V}}_N = \left(\frac{\widetilde{\mathcal{A}}_{m}^{(N)}}{\widetilde{\mathcal{B}}^{(N)}_{m-1}} , \frac{\widetilde{\mathcal{B}}^{(N)}_{m}}{N\widetilde{\mathcal{A}}_m^{(N)}}\quad, \text{ for } m\in \mathbb{Z}\right)\ .
\]
Explicitly, one obtains
\begin{align*}
\widetilde{\mathcal{V}}_5 &= \left(\tfrac12\left(1-\tfrac1{\sqrt5}\right),\ldots,\tfrac27,\tfrac3{10},\tfrac13,\tfrac25,\tfrac12,\tfrac35,\tfrac23, \tfrac7{10},\tfrac5{7},\ldots,\tfrac12\left(1+\tfrac1{\sqrt5}\right) \right) \\
\widetilde{\mathcal{V}}_6 &= \left(\tfrac12\left(1-\tfrac1{\sqrt3}\right),\ldots,\tfrac{3}{14},\tfrac29,\tfrac14,\tfrac13,\tfrac12,\tfrac23,\tfrac34, \tfrac79,\tfrac{11}{14},\ldots,\tfrac12\left(1+\tfrac1{\sqrt3}\right) \right)
\end{align*}
We define the polygon $\mathcal{M}^{(N)}$, for $N=5,6$, from the vertices $\mathcal{V}_N\cup \widetilde{\mathcal{V}}_N$ with edges formed by connecting adjacent points in the sequence of vertices. 


\subsection{Roots and Lorentzian lattices}

A vector $X\in \mathbb{R}^{2,1}$ can be represented by a $2\times 2$ symmetric matrix 
\[
X = \begin{pmatrix} t+y & x \\ x & t-y \end{pmatrix}\ , \quad x,y,t \in \mathbb{R}
\]
with norm $(X,X):=-2\det X= 2(-t^2 + x^2 +y^2)$. The inner product of two vectors $X$ and $Y$ is given by
\[
(X,Y) = - \det(Y)\ \text{Tr}(X Y^{-1})\ .
\]
Define the positive (future) light-cone by the set of time-like vectors
\begin{equation}
V^+ = \left(X\in \mathbb{R}^{2,1}| \det(X)>0, \text{Tr}(X)>0\right) \ .
\end{equation}
Further imposing $\det(X)=+1$ gives a hyperboloid which can be mapped to the upper-half plane. Figures \ref{Mfive} and \ref{Msix} are the hyperbolic polygons drawn in the upper-half plane model for the hyperboloid.
Let $\mathcal{H}_\alpha$ denote the hyperplane $(X,\alpha)=0$ for some space-like   $\alpha\in \mathbb{R}^{2,1}$. The hyperboloid given by $\det X=1$ always intersects the hyperplane $\mathcal{H}_\alpha$. The  edge of the hyperbolic polygon appear in this fashion.

\subsubsection{Roots from polygons}

Following the observation of Cheng and Verlinde\cite{Cheng:2008fc}, we associate a simple root to each wall of the hyperbolic polygons $\mathcal{M}^{(N)}$. 
To each edge of the polygon with vertices $(\frac{b}a,\frac{d}c)$ we associate a root (represented as a vector in $\mathbb{R}^{2,1}$) using the mapping
\begin{equation}
\left (\tfrac{b}a, \tfrac{d}c\right) \longleftrightarrow \alpha = \begin{pmatrix} 2bd & ad + bc \\ ad + bc & 2 ac \end{pmatrix}\ ,
\end{equation}
with norm $(\alpha,\alpha) = 2 (ad -bc)^2=2$. For example, the following two edges appear for all $\mathcal{M}^{(N)}$:
\begin{equation}
(\tfrac{-1}{0},\tfrac01) \longleftrightarrow \alpha_0 =\begin{pmatrix} 0 & -1 \\ -1 & 0 \end{pmatrix}\quad,\quad
(\tfrac11,\tfrac10) \longleftrightarrow \beta_0 =\begin{pmatrix} 2 & 1 \\ 1 & 0 \end{pmatrix}\ ,
\end{equation}
where we have represented  $i\infty$  by $\frac{-1}{0}$ for the first case and by $\tfrac10$ for the second case. Let
\begin{equation}
(\tfrac01,\tfrac1N) \longleftrightarrow \beta_{-1} =\begin{pmatrix} 0 & 1 \\ 1 & 2N \end{pmatrix}\ ,\ 
(\tfrac{N-1}N,\tfrac11) \longleftrightarrow \alpha_{-1} =\begin{pmatrix} 2(N-1) & 2N-1 \\ 2N-1 & 2N \end{pmatrix}
\end{equation}
denote the edges adjacent to the ones given above. The other roots are obtained recursively as follows:
\begin{equation}\label{gammaNdef}
\alpha_{m+1} = \gamma^{(N)} \cdot \alpha_m \cdot \left(\gamma^{(N)}\right)^T \quad,\quad
\beta_{m-1} = \gamma^{(N)} \cdot \beta_m\cdot  \left(\gamma^{(N)}\right)^T \ ,
\end{equation}
where $\gamma^{(N)}=\left(\begin{matrix}
                1 & -1 \\  N & 1-N
               \end{matrix}  \right)$.
Combining all the edges that appear from the vertices in $\mathcal{V}_N$, we define
\begin{equation}
\mathbf{X}_N = (\ldots, x_{-1},x_0,x_1,x_2,\ldots) := (\ldots, \beta_{-1},\alpha_0,\beta_0, \alpha_{-1},\ldots )\ ,
\end{equation}
i.e., $x_{2m}=\alpha_{-m}$ and $x_{2m+1}=\beta_m$ for $m\in \mathbb{Z}$. The Cartan matrix for the real simple roots is given by
\begin{equation}\label{CartanMatrix}
A^{(N)}:=(a_{nm})\quad\text{where}\quad a_{nm} = (x_n,x_m)=2 - \tfrac{4}{N-4} (\lambda_N^{n-m} + \lambda_N^{m-n}-2)\ .
\end{equation}
\textbf{Remarks:} (i) For $N<4$, the matrices are finite dimensional and the values $n$ and $m$ are defined modulo 3,4,6 (respectively) for $N=1,2,3$(respectively). (ii) When $N=4$, the matrix is determined by a limiting procedure leading to $a_{nm}=2-4(n-m)^2$ where $n,m\in \mathbb{Z}$. (iii) For $N=5,6$, again $n,m\in \mathbb{Z}$ but these corresponds to half the real simple roots.

For $N=5,6$ the other sets of roots can be generated by defining\cite{ Sen:2007vb,Krishna:2010gc}
\begin{equation}
\widetilde{\mathbf{X}}_N = (\ldots, \widetilde{x}_{-1},\widetilde{x}_0,\widetilde{x}_1,\widetilde{x}_2,\ldots)=(\ldots, \widetilde{\beta}_{-1},\widetilde{\alpha}_0,\widetilde{\beta_0}, \widetilde{\alpha}_{-1},\ldots )\ ,\ 
\end{equation}
where $\widetilde{x}_{-m} = \sigma^{(N)} \cdot x_m \cdot (\sigma^{(N)})^T$ for all $m\in \mathbb{Z}$
and
\[
\sigma^{(5)}=\begin{pmatrix}
2 & -1 \\ 5 & -2
\end{pmatrix}\quad,\quad
\sigma^{(6)}=\frac1{\sqrt2} \begin{pmatrix}
2 & -1 \\ 6 & -2 
\end{pmatrix}\quad.
\]
Note that $\sigma^{(5)}\in \Gamma_0(5)$ while $\sigma^{(6)}\in PSL(2,\mathbb{R})$.
For $N=5$, for instance, one finds
\[
\left(\tfrac25,\tfrac12\right)\longleftrightarrow \widetilde{\alpha}_0 =\begin{pmatrix} 4 & 9 \\ 9 & 20 \end{pmatrix} \quad ,\quad
\left(\tfrac12,\tfrac35\right)\longleftrightarrow \widetilde{\beta}_{0} = \begin{pmatrix} 6 & 11 \\ 11 &20 \end{pmatrix}\quad.
\]
Similarly, for $N=6$, one finds
\[
\left(\tfrac23,\tfrac12\right)\longleftrightarrow\tilde{\alpha}_0 =\begin{pmatrix} 2 & 5 \\ 5 & 12 \end{pmatrix} \quad ,\quad
\left(\tfrac12,\tfrac23\right)\longleftrightarrow\tilde{\beta}_{0} = \begin{pmatrix} 4 & 7 \\ 7 &12  \end{pmatrix}\quad.
\]
The following inner products hold
\begin{equation}\label{innerproducts}
(x_n,x_m)=(\widetilde{x}_m, \widetilde{x}_n)= (\widetilde{x}_m,x_n)+ c_N\ ,
\end{equation}
with $c_5=20$ and $c_6=12$.

Let $s_{x_m}$ denote the elementary (Weyl) reflection generated by the root $x_m\in \mathbf{X}_N$. 
\[
s_{x_m}:\quad \alpha \rightarrow \alpha - 2\ \frac{(\alpha, x_m) }{(x_m,x_m)}\ x_m\ ,
\]
for $\alpha\in \mathbb{R}^{2,1}$ and let $W_+$ be the group generated by these Weyl reflections. Similarly, let $s_{\tilde{x}_m}$ denote the elementary (Weyl) reflection generated by the root $\tilde{x}_m\in \mathbf{\tilde{X}}_N$.  Let $W$ be the group generated by all elementary reflections generated by  $s_{x_m}$ and $s_{\tilde{x}_m}$ for all $m\in \mathbb{Z}$.

\subsubsection{Symmetries of the hyperbolic polygon}

There is a dihedral symmetry with generators $a$ and $b$ satisfying 
\begin{equation}\label{dihedralrels}
a^2=b^2=1\text{ and }(ab)^{m(N)}=1\ ,
\end{equation}
with  $m(N)=3,2,3$ for $N=1,2,3$ respectively and $m(N)=\infty$ for $N>3$. The generator $a=\delta$ acts on the roots as follows:
\begin{equation}\label{deltadef}
\delta \cdot \alpha \cdot \delta^T \quad \textrm{with } \delta =\begin{pmatrix} -1 & 1 \\ 0 & 1\end{pmatrix}\ ,
\end{equation}
and $\alpha$ is any root. Note that $\delta^2=-1$ which has trivial action on the roots. Similarly, $b=\gamma^{(N)}\cdot \delta$, where the action of $\gamma^{(N)}$ on the roots is given in Eq. \eqref{gammaNdef}. Again the matrix $b^2=-1$ which has trivial action on the roots.

Let Dih$(\mathcal{M}^{(N)}):=\langle a, b\rangle$. For $N\leq 4$, this dihedral group is the symmetry group of the hyperbolic polygon, $\mathcal{M}^{(N)}$ which we denote by Sym($\mathcal{M}^{(N)})$.
For $N=5$ and $6$, there is an  additional generator, $\sigma^{(N)}$, that has the following action on the generators of the dihedral group:
\begin{equation}\label{addrels}
\sigma^{(N)} \cdot a \cdot (\sigma^{(N)})^{-1} = b \quad, \quad \sigma^{(N)} \cdot b \cdot (\sigma^{(N)})^{-1} = a\ .
\end{equation}
On the roots, the generators of the dihedral group act  as:
\begin{align}
a:& \quad \alpha_m \leftrightarrow \beta_{m~~} \quad,\quad \widetilde{\alpha}_{m} \leftrightarrow \widetilde{\beta}_{m} \\
b: &\quad \alpha_m \leftrightarrow \beta_{m-1} \quad,\quad \widetilde{\alpha}_m \leftrightarrow \widetilde{\beta}_{m-1} 
\end{align}
The generator $\sigma^{(N)}$ is such that it acts as an order two element on the roots even though its matrix realisation is such that $(\sigma^{(N)})^4=1$. Thus, for $N=5$ and $6$ we have 
\begin{equation} 
\text{Sym}(\mathcal{M}^{(N)}) = \langle a, b, \sigma^{(N)}\rangle \ ,
\end{equation}
where we assume $(\sigma^{(N)})^2=1$.  Note that for $N\leq 4$, Sym$(\mathcal{M}^{(N)})=\text{Dih}(\mathcal{M}^{(N)})$.
Following, Cheng and Dabholkar\cite{Cheng:2008kt}, we call the group $(W_+\rtimes \text{Dih}(\mathcal{M}_N))$ as the \textit{extended S-duality} group.


\subsubsection{Root Lattices with Weyl vector}

We have seen that the hyperbolic polygons $\mathcal{M}^{(N)}$ are determined by the vertices $\mathcal{V}_N$ and edges (represented by the $SL(2,\mathbb{Z})$ matrices) $\mathbf{X}_N$ for $N\leq 4$. For $N=5,6$, the vertices are $\mathcal{V}_N\cup \widetilde{\mathcal{V}}_N$ and the edges are given by $\mathbf{X}_N \cup \widetilde{\mathbf{X}}_N$.

Define the lattice in $\mathbb{R}^{2,1}$ for $N\leq 4$
\begin{equation}
\mathcal{L}^{(N)} = \oplus_{x_m \in \mathbf{X}_N} \mathbb{Z}\, x_m\ .
\end{equation}
and for $N=5,6$
\begin{equation}
\mathcal{L}^{(N)} 
= \Big(\oplus_{x_m \in \mathbf{X}_N} \mathbb{Z}\, x_m\Big) \oplus \left( \oplus_{\widetilde{x}_m \in \widetilde{\mathbf{X}}_N} \mathbb{Z}\, \widetilde{x}_m\right)\ .
\end{equation}
This is a rank-three Lorentzian lattice with a Weyl vector 
\begin{equation}
\varrho^{(N)}=\begin{pmatrix} 1/N & 1/2 \\ 1/2 & 1\end{pmatrix}\text{ with } \big(\varrho^{(N)},\varrho^{(N)}\big)=\frac12 -\frac2N\ .
\end{equation}
The Weyl vector has the following properties:
\begin{enumerate}
\item The norm of $\varrho^{(N)}$ is $\left(\frac{N-4}{2N}\right)$. Thus, the norm is time-like for $N<4$, light-like for $N=4$ and space-like for $N=5,6$.
\item The inner product of the Weyl vector with real simple roots are:
\[
 (\varrho^{(N)}, x_m)=-1\quad  \forall\ x_m \in \mathbf{X}_N\ .
\]
and for $N=5,6$, additionally one has
\[
 (\varrho^{(N)}, \widetilde{x}_m)=+1\quad  \forall\ \widetilde{x}_m \in \widetilde{\mathbf{X}}_N\ .
\]
\item 
The generators of  Sym$(\mathcal{M}_N)$ act on the Weyl vector as follows:
\[
a: \varrho^{(N)}\rightarrow \varrho^{(N)} \ ,\quad
b: \varrho^{(N)}\rightarrow \varrho^{(N)}, \  \text{ and } \quad \sigma^{(N)}:\varrho^{(N)}\rightarrow -\varrho^{(N)}\ .
\]
Thus Dih$(\mathcal{M}_N)$ preserves the Weyl vector.
\end{enumerate}

The rank-three hyperbolic root lattices $\mathcal{L}^{(N)}$ with a Weyl vector $\varrho^{(N)}$ and hyperbolic polygons $\mathcal{M}^{(N)}$ that we obtain fit in with Nikulin's classification of hyperbolic root systems of rank three\cite{Nikulin:2000}. In particular, the lattices for $N<4$ are with Weyl vector of \textit{elliptic} type, the lattice for $N=4$ is with Weyl vector of \textit{parabolic} type and the lattices for $N=5,6$ are with Weyl vector of \textit{hyperbolic} type.


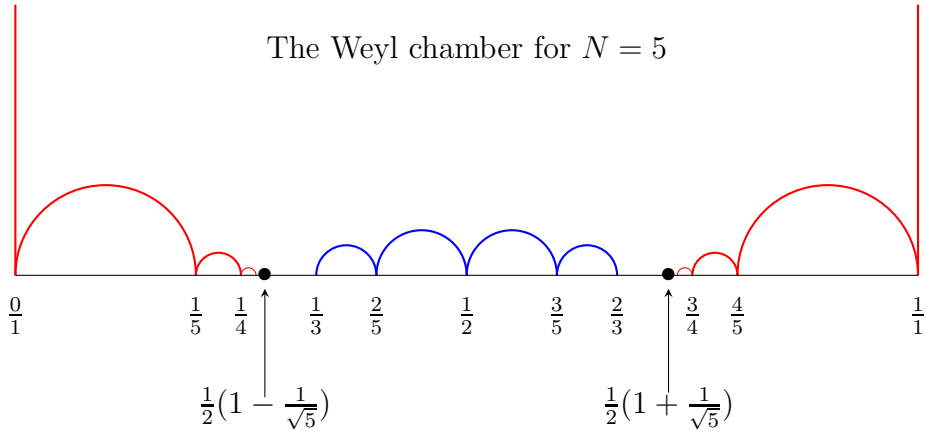
\begin{figure}[htb]
\begin{tikzpicture}[scale=12]
 \draw[thick,color=red] (0,0) arc(180:0:1/10) (1/5,0);
 \draw[thick,color=red] (1/5,0) arc(180:0:1/40) (1/4,0);
  \draw[color=red] (1/4,0) arc(180:0:1/120) (4/15,0);
 \draw[thick,color=red] (4/5,0) arc(180:0:1/10) (1,0);
 \draw[thick,color=red] (3/4,0) arc(180:0:1/40) (4/5,0);
   \draw[color=red] (11/15,0) arc(180:0:1/120) (3/4,0);
 \draw[thick,color=blue] (2/5,0) arc(180:0:1/20) (1/2,0);
 \draw[thick,color=blue] (1/3,0) arc(180:0:1/30) (2/5,0);
 \draw[thick,color=blue] (1/2,0) arc(180:0:1/20) (3/5,0);
 \draw[thick,color=blue] (3/5,0) arc(180:0:1/30) (2/3,0);
 \draw[thick,color=red] (0,0) -- (0,0.3);
  \draw[thick,color=red] (1,0) -- (1,0.3);
  \draw (0,0) -- (1,0);
  \node [label=below:$\tfrac01$] (X) at (0,0) {};  
    \node  at (1/2,0.25){The Weyl chamber for $N=5$};
  \node [label=below:$\tfrac11$]  at (1,0) {};
  \node [label=below:$\tfrac12$] (X) at (1/2,0) {};
   \node [label=below:$\tfrac13$] at (1/3,0) {};
  \node [label=below:$\tfrac14$] at (1/4,0) {};
  \node [label=below:$\tfrac15$] at (1/5,0) {};
   \node [label=below:$\tfrac23$] at (2/3,0) {};
  \node [label=below:$\tfrac25$] at (2/5,0) {};
    \node [label=below:$\tfrac34$] at (3/4,0) {};
  \node [label=below:$\tfrac35$] at (3/5,0) {};
  \node [label=below:$\tfrac45$] at (4/5,0) {};
  \draw (0.7236,0) node{$\bullet$};
  \node [label=below:$\tfrac12(1+\tfrac1{\sqrt5})$] at (0.7236,-0.1) {};
  \draw[-stealth] (0.7236,-0.13) -> (0.7236,-0.015);
 \draw (0.2764,0) node{$\bullet$};
 \node [label=below:$\tfrac12(1-\tfrac1{\sqrt5})$] at (0.2764,-0.1) {};
  \draw[-stealth] (0.2764,-0.135) -> (0.2764,-0.015);
\end{tikzpicture}
 \caption{The Weyl chamber with some of the (infinite) roots of the hyperbolic polygon $\mathcal{M}^{(5)}$. The two dark circles indicate the limit points of the infinite sets of roots.}\label{Mfive}
 \end{figure}
 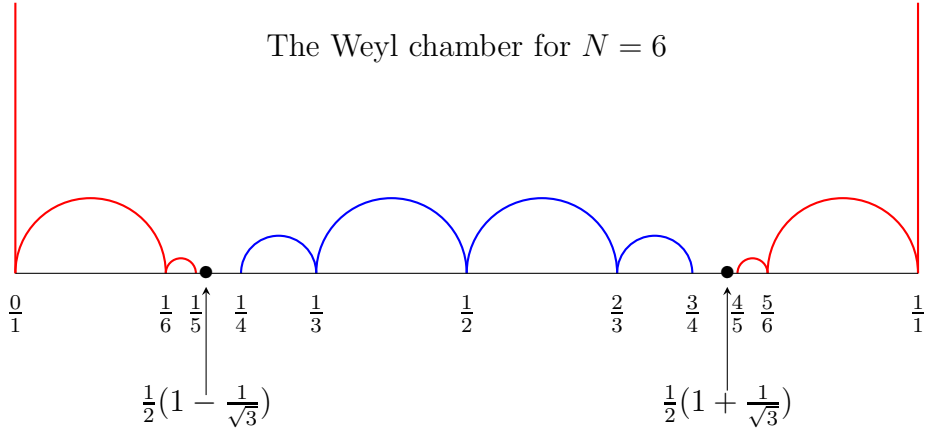
\begin{figure}[htb]
\begin{tikzpicture}[scale=12]
 \draw[thick,color=red] (0,0) arc(180:0:1/12) (1/6,0);
 \draw[thick,color=red] (1/6,0) arc(180:0:1/60) (1/5,0);
 \draw[thick,color=red] (5/6,0) arc(180:0:1/12) (1,0);
 \draw[thick,color=red] (4/5,0) arc(180:0:1/60) (5/6,0);
 \draw[thick,color=blue] (1/3,0) arc(180:0:1/12) (1/2,0);
 \draw[thick,color=blue] (1/4,0) arc(180:0:1/24) (1/3,0);
 \draw[thick,color=blue] (1/2,0) arc(180:0:1/12) (2/3,0);
 \draw[thick,color=blue] (2/3,0) arc(180:0:1/24) (3/4,0);
 \draw[thick,color=red] (0,0) -- (0,0.3);
  \draw[thick,color=red] (1,0) -- (1,0.3);
  \draw (0,0) -- (1,0);
  \node [label=below:$\tfrac01$] (X) at (0,0) {};  
  \node  at (1/2,0.25){The Weyl chamber for $N=6$};
  \node [label=below:$\tfrac11$]  at (1,0) {};
  \node [label=below:$\tfrac12$] (X) at (1/2,0) {};
   \node [label=below:$\tfrac13$] at (1/3,0) {};
  \node [label=below:$\tfrac16$] at (1/6,0) {};
  \node [label=below:$\tfrac15$] at (1/5,0) {};
   \node [label=below:$\tfrac23$] at (2/3,0) {};
  \node [label=below:$\tfrac14$] at (1/4,0) {};
    \node [label=below:$\tfrac56$] at (5/6,0) {};
  \node [label=below:$\tfrac34$] at (3/4,0) {};
  \node [label=below:$\tfrac45$] at (4/5,0) {};
  \draw (0.7887,0) node{$\bullet$};
  \node [label=below:$\tfrac12(1+\tfrac1{\sqrt3})$] at (0.7887,-0.1) {};
  \draw[-stealth] (0.7887,-0.13) -> (0.7887,-0.015);
 \draw (0.2113,0) node{$\bullet$};
 \node [label=below:$\tfrac12(1-\tfrac1{\sqrt3})$] at (0.2113,-0.1) {};
  \draw[-stealth] (0.2113,-0.135) -> (0.2113,-0.015);
\end{tikzpicture}
 \caption{The Weyl chamber with some of the (infinite) roots of the hyperbolic polygon $\mathcal{M}^{(6)}$. The two dark circles indicate the limit points.}\label{Msix}
\end{figure}

\section{Siegel Modular Forms}\label{SecSiegel}

In the previous section, we constructed a special family of root lattices that arose from the study of walls of marginal stability in CHL $\mathbb{Z}_N$ orbifolds. In this section, we will consider multiplicative eta products and Siegel modular forms that arise as the generating function of half and quarter BPS states in the same CHL $\mathbb{Z}_N$ orbifolds. Let $g$ denote the discrete symmetry of order $N$ that leads to the CHL orbifold. 

\subsection{Mathieu moonshine}

Mathieu moonshine refers to the connection between the sporadic simple group $M_{24}$ and a variety of modular forms. The first such moonshine mapped conjugacy classes of $M_{24}$ and multiplicative eta products\cite{Dummit:1985,Mason:1985}. Conjugacy classes of $M_{24}$ can be represented by cycle shapes as it is a sub-group of the symmetric group on 24 letter, $S_{24}$. Let $\rho=1^{a_1}\cdots N^{a_N}$ be a conjugacy class of an element, $g$, of order $N$ in $M_{24}$. Then, the eta product associated with this conjugacy class is given by the product 
\begin{equation}\label{metaprod}
\eta_\rho(\tau):= \prod_{j=1}^N \eta(j \tau)^{a_j}\ ,
\end{equation}
is a multiplicative eta product. In \cite{Govindarajan:2009qt}, it was shown that the inverse of this eta product naturally appeared as a refined generating function of  $\tfrac12$-BPS states in type II string theory compactified on $K3\times T^2$. These generating functions arise as four-dimensional indices given by a helicity trace $B_4$ which receive contributions from states that break half the spacetime supersymmetry\cite{Gregori:1997hi}. In \cite{Sen:2009md}, Sen introduced a refined index, $B_4^g$, by the insertion of a discrete symmetry $g$ in the helicity trace -- we shall call this twining by the symmetry $g$. The observation of \cite{Govindarajan:2009qt} is that when $g$ is a symplectic automorphism of $K3$, then the inverse of the multiplicative eta product is the $g$-twined helicity trace. 

Another helicity trace, $B_6$ and its refinement $B_6^g$ similarly receives  contributions from $\tfrac14$-BPS states and is the inverse of a genus-two Siegel modular form\cite{Sen:2009md}. In \cite{Govindarajan:2009qt}, it was shown that this genus-two Siegel modular form (associated with the symplectic automorphism $g$) is given by the additive lift  of a Jacobi form constructed from multiplicative eta product for the same element $g$. The Jacobi form is given by
\begin{equation}
\phi^\rho(\tau,z) = \frac{\vartheta_1(\tau,z)^2}{\eta(\tau)^6} \times \eta_\rho(\tau)\ .
\end{equation} 

These Siegel modular forms, $\Phi^\rho(\mathbf{Z})$, have a second construction that leads to a product formula generated by a weight zero, index one Jacobi form that is given by the $g$-twined elliptic genus of $K3$. This generalizes the observation of Eguchi, Ooguri and Tachikawa that the dimensions of irreps of $M_{24}$ appeared in the coefficients of the expansion of elliptic genus of K3 in terms of characters of the two-dimensional $\mathcal{N}=4$ superconformal algebra\cite{Eguchi:2010ej}. The construction of the Jacobi forms for all conjugacy classes of $M_{24}$ was completed in \cite{Gaberdiel:2010ca,Eguchi:2010fg}. We denote them by $\psi^\rho(\tau,z)$. The product formulae for all these Siegel modular forms were obtained in \cite{Govindarajan:2011em,Cheng:2012uy}. The proof of modularity of the product formula for all conjugacy classes was only completely recently by one of us in\cite{Sutapa2019a} (see also \cite{Raum:2012,Govindarajan:2018ted}). The additive lift does not exist for all conjugacy classes. However, the Jacobi form constructed out of the multiplicative eta product i.e., $\phi^\rho(\tau,z)$, appears as the ``zeroth Fourier-Jacobi coefficient".

\subsection{Generalized Mathieu moonshine}

In the context of Monstrous moonshine, Norton introduced the notion of generalized moonshine\cite[see appendix by Norton]{Mason:1987}. In our context, this implies that given two commuting elements of $(g,h)\in M_{24}$, one associates a modular form that generalizes the eta product, Jacobi forms and Siegel modular forms that appear for each conjugacy class of $M_{24}$. We refer to this as generalized Mathieu moonshine.
In ref. \cite{Mason1990}, to a subset of such commuting pairs $(g,h)$ that he calls rational, Mason constructs eta products  by associating a Frame shape to each pair. We represent these eta products by $\eta^{[g,h]}(\tau)$ -- this is the eta product corresponding to \textit{twining} by the element $g$ and \textit{twisting} by the element $h$  (see appendix \ref{Secmoonshineproduct} for notation). In particular, note that Mason has  a different notation from ours. The same eta product is written as $\eta^{[h,g]}(\tau)$.). The formula for the general case is complicated and hence we give them for the cases of interest. First, we consider $g$ has order $N$ and  let $\rho=1^{a_1}\cdots N^{a_N}$ be the cycle  shape associated with it. Then, for $(a,b)=1$, and $\gamma=\left(\begin{smallmatrix} * & * \\ b & a\end{smallmatrix}\right)\in SL(2,\BZ)$, 
\begin{equation}\label{genetaproduct}
\eta^{[g^a,g^b]}(\tau) =   (\text{constant})\ \eta_\rho(\tau)|_k \gamma\ .
\end{equation}
where the constant is determined on a case by case basis.
In this notation, $\eta^{[g,1]}(\tau)$ is the multiplicative eta product $\eta_\rho(\tau)$ associated with the $M_{24}$ conjugacy class, $\rho$ of $g$ as defined in Eq. \eqref{metaprod}. In the examples involving umbral moonshine that we consider,
\begin{equation}\label{genetaproductb}
\eta^{[g,h]}(\tau) :=    \eta_\rho(\tfrac\tau{N} ) \ . 
\end{equation}
 The explicit eta product multiplied by $\vartheta_1(\tau,z)^2/\eta(\tau)^6$ appears in the entries with lambency $\leq 5$ in column three of Table \ref{umbraltable}. The cycle shape, $\rho$, is also given as column two of Table \ref{periodictable}.

Similarly, we represent the elliptic genus of K3 given by twining by the element $g$ and twisted by the element $h$ by $\psi_{0,1}^{[g,h]}(\tau,z)$\cite{Gaberdiel:2012gf}. For instance, $\psi_{0,1}^{[g,1]}(\tau,z)$ corresponds to the Mathieu moonshine Jacobi form, $\psi^\rho(\tau,z)$,  for the $M_{24}$ conjugacy class of $g$.  For some pairs, the elliptic genus vanishes. In some of those cases, one finds a Jacobi form of higher index associated with this pair as originally claimed in \cite{Govindarajan:2010fu}. These higher index Jacobi forms turn out to be related to umbral moonshine, a generalization of Mathieu moonshine\cite{Cheng:2012tq}. To each commuting pair, one naturally obtains a potential Siegel modular form that we denote by $\Phi_k^{[g,h]}(\textbf{Z})$ given by a product formula given in Eq. \eqref{moonshineproduct}. When possible, we combine this with two other constructions to prove modularity of the product formula for three families of examples.


\subsection{The $\mathbb{Z}_N$ CHL orbifolds}

We have seen that $\mathbb{Z}_N$ CHL orbifolds of the heterotic string compactified on a six-torus are associated to rank three hyperbolic lattices for $N\leq 6$. The generating function of quarter BPS states in these theories which we will denote by $\Phi^{[1,g]}(\mathbf{Z})$ is a Siegel modular form of a level $N$ subgroup of $\mathbf{\Gamma}_1:=Sp(2,\BZ).$\footnote{$\mathbf{Z}=\left(\begin{smallmatrix} \tau & z \\ z & \tau'\end{smallmatrix}\right)$ is a point in the Siegel upper half space. We also set $q=\exp(2\pi i \tau)$, $r=\exp(2\pi iz)$ and $s=\exp(2\pi i \tau')$.} The generator of the $\BZ_N$ orbifold $g$ is identified with a symplectic automorphism of $K3$ in the dual type IIA picture as a symmetric orbifold of type IIA compactified on $K3\times T^2$. Thus $g$ can be identified with an element of $M_{23}\subset M_{24}$ and the order of $g$ uniquely fixes its conjugacy class or equivalently its cycle shape. In all these cases, the square-root of the $\Phi^{[1,g]}(\mathbf{Z})$ makes sense and we will write
\begin{equation}
\Phi^{[1,g]}(\mathbf{Z}) = \left(\Delta^{[1,h]}(\mathbf{Z})\right)^2\ ,
\end{equation}
where $h$ is identified with an element of $L_2(11)_B\subset M_{12}$ for $N\neq 4$. There are two $L_2(11)$ sub-groups of $M_{12}$ which are inequivalent under conjugation in $M_{12}$\cite{Conway:1979}. In the notation introduced in \cite{Govindarajan:2018ted}, the subgroup $L_2(11)_B$ is $L_2(11)_B\subset M_{11}\subset M_{12}$. $L_2(11)_A$, on the other hand, is a maximal subgroup of $M_{12}$.
\begin{table}[hbt]
\centering
\newcommand\T{\rule{0pt}{2.6ex}}
\begin{tabular}{c|cccccc} \hline
$N$ & 1 & 2& 3 & 4 & 5 & 6 \T \\[2pt] \hline 
$[g]\in M_{23}$ & $1^{24}$ &  $1^82^8$ & $1^63^6$ & $1^42^24^4$ & $1^45^4$ & $1^22^23^26^2$  \T \\[3pt] \hline
$[h]\in L_2(11)_B$ & $1^{12}$ & $1^42^4$ & $1^33^3$ & -- & $1^25^2$ & $1^12^13^16^1$ \T \\[3pt] \hline
\end{tabular}
\caption{Table of cycle shapes.}\label{cycleshapes}
\end{table}

These modular forms are constructed in two different ways:
\begin{enumerate}
\item As a product formula that follows from moonshine: $M_{24}$ for $\Phi^{[1,g]}(\mathbf{Z})$ and $L_2(11)_B$ for $\Delta^{[1,g]}(\mathbf{Z})$.
\item Directly as a lift that is implied by the S-transform of an additive lift. 
\end{enumerate}
The equality of the two constructions proves modularity of the product formula given by moonshine. 

\subsubsection*{Product Formulae}
Define the 
index one Jacobi form
\begin{equation}
\phi_{k,1}^{[g^b,g^d]}(\tau,z) = \frac{\theta_1(\tau,z)^2}{\eta(\tau)^6}\  \eta^{[g^b,g^d]}(\tau)\ ,
\end{equation}
and let $\psi^{[g^{b},g^d]}_{0,1}(\tau,z)$ be the $g^d$-twisted elliptic genus of $K3$ twined by the element $g^{b}$ defined in Eq. \eqref{twistedtwinedgenera}. Consider the Fourier expansion
\begin{equation}\label{Fouriercoeffs}
\psi^{[g^{b},g^d]}_{0,1}(\tau,z)=\sum_{n\in\mathbb{Z}, n\ge0}\sum_{\ell \in \mathbb{Z}} c^{[b,d]}(n,\ell)\ q^{\frac{n}{N}} r^\ell\ ,
\end{equation}
where  $g$ is of order $N$.  Note that both $b$ and $d$ are only defined modulo $N$ and that when the order of $(g^d)<N$, several coefficients might be zero. This enables us to write simpler formulae.  Explicit formulae for these are available for $N\in \{1,\ldots,5\}$ in the literature\cite{David:2006yn,David:2006ji,David:2006ud} and we give the formulae for $N=6$ in section \ref{subsecothercusps}.  The discrete Fourier-transform of the coefficient $c^{[b,d]}(n,\ell)$ is given by 
\begin{align}\label{DFTcoeff}
c^{[b,d]}(n,\ell)=\sum_{\alpha =0}^{N-1} \hat{c}^{[\alpha,d]}(n,\ell)\exp\left(\frac{2\pi i \alpha b}{N}\right)\ .
\end{align}
The product formula for $\Phi_k^{[1,g]}(\mathbf{Z})$ is then given by setting $x=0$ and $y=1$ in Eq. \eqref{moonshineproduct}, we obtain
\begin{align}
\Phi_k^{[1,g]}(\mathbf{Z})=s\phi_{k,1}^{[1,g]}(\tau,z) \times\prod_{m=1}^\infty\prod_{\alpha=0}^{N-1}\prod_{\substack{n\in \mathbb{Z}-\frac{\alpha}{N}\\ n\ge0}}\prod_{\ell\in \mathbb{Z}}(1-q^{n} r^
\ell s^m)^{ {\hat{c}}^{[\alpha,m]}(nmN,\ell)}\ .
\end{align}
The above formula follows from a computation starting from the twisted Hecke operator defined in Eq. \eqref{TwistedHeckeOperator} and the formula for the Siegel modular form given in Eq. \eqref{PhiGeneral} applied to the specific case at hand. For the square root to make sense, one needs (i) $c^{[\alpha,m]}(nm,\ell)$ to be even and (ii) the multiplicative eta products appear with even powers. This is true in all the six cases and one obtains the following product formula.
\begin{align}
\Delta_{k/2}^{[1,h]}(\mathbf{Z})=s^{1/2} \sqrt{\phi_{k,1}^{[1,g]}(\tau,z) } \times\prod_{m=1}^\infty\prod_{\alpha=0}^{N-1}\prod_{\substack{n\in \mathbb{Z}-\frac{\alpha}{N}\\ n\ge0}}\prod_{\ell\in \mathbb{Z}}(1-q^{n} r^
\ell s^m)^{\tfrac12 {\hat{c}}^{[\alpha,m]}(nmN,\ell)}\ .
\end{align}

\subsubsection*{Additive Lift}

The additive lift of  Cl\'ery and Gritsenko\cite{Gritsenko:2008} begins with a a Jacobi form of weight $k$ and index $t\in \tfrac12 \mathbb{Z}$ for the sub-group $\Gamma_0(Nq,q)\in PSL(2,\BZ)$. For each $m>0$ such that $(m,q)=1$, they define a Hecke-like operator $T_-(m)$ defined by
\begin{equation}\label{CGHecke}
T_-(m) := \sum_{\substack{ad=m\\ (a,N)=1\\a>0}} \Gamma_1(Nq,q)\ \sigma_a \begin{pmatrix} a & bq \\ 0 & d \end{pmatrix}\ ,
\end{equation}
where $\sigma_a =\left(\begin{smallmatrix} a^{-1} & 0 \\ 0 & a \end{smallmatrix}\right)\text{ mod }Nq$. The Hecke-like operator maps the given Jacobi form to another Jacobi form of weight $k$ and index $mt$.
The additive lift of a Jacobi form $\phi_{k,t}(\tau,z)$ is then given by
\[
 \Delta_k^{[g,1]}(\mathbf{Z}) = \sum_{m=1 \text{ mod }q} s^{mt}\ \phi\Big|_{k,t}\ T_-(m) (\tau,z)
\]
For all our examples, the index of the Jacobi form $t=1/2$ and $q=2$. The  $ \Delta_k^{[g,1]}(\mathbf{Z}) $ are modular forms with character of the level $N$ subgroup $\Gamma_1(N,1,N,1)$  of $\Gamma_1=Sp(2,\BZ)$ as defined in appendix \ref{modularstuff}. 

Recall that $S\cdot  \Gamma_0(Nq,q)\cdot  S^{-1}=\Gamma_0(q,Nq)$. So the additive lift for $\Gamma_0(q,Nq)$ is equivalent to the $S$-transform of the above lift for $\Gamma_0(q,Nq)$. Thus, one has the additive lift leading to a Siegel modular form.
\begin{align}
\Delta^{[1,g]}_k(Z) &= \sum_{m=1\text{ mod } q} s^{mt} \ \widetilde{\phi}_{k,mt}(\tau,z) \nonumber  \\
&:=\sqrt{\tfrac{C_\rho}{(-i)^{2k+2}}} \sum_{m=1\text{ mod } q} s^{mt} \ \left(\phi_{k,t} \Big|_k T_-(m)\cdot S\right)(\tau,z)\ .
\end{align}
The $\Delta^{[1,g]}_k(Z)$ are Siegel modular forms of the with character of the level $N$ subgroup $(\tilde{S}\cdot \Gamma_1(N,1,1,1)\cdot\tilde{S}^{-1})$  of $\Gamma_1$, where
\begin{equation*}
\tilde{S} = \left(\begin{smallmatrix} 0 & 0 & -1 & 0 \\
 0 & 1 & 0 & 0 \\
 1 & 0 & 0 & 0 \\
 0 & 0 & 0 & 1 
 \end{smallmatrix}\right)\ ,
\end{equation*}
is the $S$-transform embedded into $\Gamma_1$. Explicitly the group consists of $Sp(2,\BZ)$ matrices of the form
\begin{equation}\label{SGroup}
\left(\begin{smallmatrix}
* & * & N * & * \\
 * & * & * & * \\
 * & * & * & * \\
 * & N * & N * & 1 
\end{smallmatrix}\right)\quad,\quad \textrm{where } *\in \BZ\ .
\end{equation}

One needs to rewrite this transformed Hecke operator $T_-(m)\cdot S$ in terms of $S \cdot T_-(m)$ and possibly other terms. This is done by constructing new representations for the coset elements similar to the one described, for instance, in \cite[appendix C]{Jatkar:2005bh}. We illustrate with a few explicit formulae focusing on the ones relevant for $N=5,6$.
\begin{enumerate}
\item The Hecke-like operator takes the simple form when $(m,N)=1$. Note that one also has $(m,q)=1$.
\begin{equation}
T_-(m)\cdot S =  \sum_{\substack{ad=m\\ a>0 \\  (a,N)=1 \\ c \text{ mod }a}} S \cdot \sigma_{a^{-1}}\cdot \begin{pmatrix} d & c\, Nq \\ 0 & a \end{pmatrix} \quad \text{ when }(m,N)=1
\end{equation}
When $(m,N)>1$, then extra terms appear for $a|N$ as illustrated by the next two examples.
\item When $N=m=5$ and $q=2$, we get
\begin{multline}
 T_-(5)\cdot S = S \cdot \begin{pmatrix} 5 & 0 \\ 0 & 1\end{pmatrix} 
+ \gamma_{2/5} \cdot \begin{pmatrix} 1 & 2 \\ 0 & 5\end{pmatrix} 
+ \gamma_{2/5} \cdot \begin{pmatrix} 1 & -2 \\ 0 & 5\end{pmatrix} \\
+ \gamma_{4/5} \cdot \begin{pmatrix} 1 & 1 \\ 0 & 5\end{pmatrix} 
+ \gamma_{4/5} \cdot \begin{pmatrix} 1 & -1 \\ 0 & 5\end{pmatrix} 
\end{multline}
where $\gamma_{r/p}=\left(\begin{smallmatrix} r & * \\ p & *  \end{smallmatrix}\right)\in PSL(2,\mathbb{Z})$ maps $i\infty$ to $r/p$. One has $\gamma_{2/5}=S T^{-2} S T^2 S$ and $\gamma_{4/5}=S T^{-1} S T^4 S$. The detailed implementation of this Hecke operator for the $\BZ_5$ CHL orbifold is described in appendix \ref{SimplifyHecke}.
\item When $N=6$ and $m=3$,
\begin{equation}
 T_-(3)\cdot S = S \cdot \begin{pmatrix} 3 & 0 \\ 0 & 1\end{pmatrix} 
+ \gamma_{2/3} \cdot \begin{pmatrix} 1 & 1 \\ 0 & 3\end{pmatrix} 
+ \gamma_{4/3} \cdot \begin{pmatrix} 1 & -1 \\ 0 & 3\end{pmatrix} 
\end{equation}
with $\gamma_{2/3}=S T^{-1} S T^2 ST$ and $\gamma_{4/3}=S T^{-1} S T^{-4} ST^{-1}$.
The detailed implementation of this Hecke operator for the $\BZ_6$ CHL orbifold is described in appendix \ref{SimplifyHecke}.

\end{enumerate}
Such formulae are important in obtaining the Fourier coefficients of the Siegel modular forms and thereby providing information on the sum side of the denominator formula.

\subsection{Expansions of Siegel modular forms about other cusps}\label{subsecothercusps}

Let $g$ be an element of $M_{23}$ of order $N$. The S-transform of the Siegel modular form $\Phi^{[g,1]}(\mathbf{Z})$ is the Siegel modular form associated with $\mathbb{Z}_N$ CHL orbfifold, i.e., $\Phi^{[1,g]}(\mathbf{Z})$. These have already been considered earlier. When the order of $g$ is prime, there are only two inequivalent cusps and so we are done. However, for composite order, there are other cusps. We focus on the case of $N=6$ where there are  two other cusps. For the rest of the section, $g$ has order $6$ and is in the conjugacy class $1^22^23^26^2$.   As we will show, these lead to two Siegel modular forms that we denote by $\Phi_2^{[g^2,g^3]}(\mathbf{Z})$ and $\Phi_2^{[g^3,g^2]}(\mathbf{Z})$.  

\subsubsection{Product formula from Mathieu moonshine}

The EOT Jacobi forms that correspond to twining by powers of $g$ are
\begin{align*}
&\psi^{[g,1]}(\tau,z)=\tfrac16\phi_{0,1}(\tau,z)-\left(\tfrac16 E_2^{(2)}(\tau)+\tfrac12E_2^{(3)}(\tau)-\tfrac52 E_2^{(6)}(\tau)\right)\ \phi_{-2,1}(\tau,z),\\
&\psi^{[g^2,1]}(\tau,z)=\tfrac12\phi_{0,1}(\tau,z)+\tfrac32 E_2^{(3)}(\tau)\ \phi_{-2,1}(\tau,z),\\
&\psi^{[g^3,1]}(\tau,z)=\tfrac23\phi_{0,1}(\tau,z)+\tfrac43 E_2^{(2)}(\tau)\ \phi_{-2,1}(\tau,z)\ .
\end{align*}
The Jacobi forms associated with generalized moonshine corresponding to twining and twisting by various powers of $g$ are then
\begin{align*}
&\psi^{[1,g]}(\tau,z)=\tfrac16\phi_{0,1}(\tau,z)+\left(\tfrac1{12} E_2^{(2)}(\tfrac\tau2)+\tfrac16E_2^{(3)}(\tfrac\tau3)-\tfrac5{12} E_2^{(6)}(\tfrac\tau{6})\right)\ \phi_{-2,1}(\tau,z),\\
&\psi^{[g^{2k},g^2]}(\tau,z)=\tfrac12\phi_{0,1}(\tau,z)-\tfrac12 E_2^{(3)}(\tfrac{\tau+k}3)\ \phi_{-2,1}(\tau,z),\\
&\psi^{[g^{2k-1},g^2]}(\tau,z)=\tfrac16\phi_{0,1}(\tau,z)-\tfrac{1}{12}\left(E_2^*(\tau)+4E_2^*(2\tau)+ E_2^*(\tfrac{\tau+k+1}3)-4 E_2^*\left(\tfrac{2\tau+2k-1}3\right)\right)\phi_{-2,1}(\tau,z),\\
&\psi^{[g^{3k\pm 1},g^3]}=\tfrac16\phi_{0,1}(\tau,z)-\tfrac{1}{12}\left(E_2^*(\tau)+E_2^*\left(\tfrac{\tau+k+1}2\right)+9 E_2^*(3\tau)-9 E_2^*\left(\tfrac{3\tau+k+1}2\right)\right)\phi_{-2,1}(\tau,z),\\
&\psi^{[g^{3k},g^3]}(\tau,z)=\tfrac23\phi_{0,1}(\tau,z)-\tfrac23 E_2^{(2)}(\tfrac{\tau+k}2)\ \phi_{-2,1}(\tau,z)
\end{align*}
The product formula for the two Siegel modular forms are given by
\begin{align}
\Phi_k^{[g^x,g^y]}(\mathbf{Z})= s\phi_{k,1}^{[g^x,g^y]} \times \prod_{m=1}^{\infty} \prod_{\alpha=0}^{N-1}\prod_{\substack{n\in\mathbb{Z}-\frac{\alpha y}{N}\\ n>0}}\prod_{\ell\in \mathbb{Z}}  (1- e^{\frac{2\pi i \alpha x}N}q^{n}r^{\ell} s^{m})^{\hat{c}^{[\alpha,my]}(nm,\ell)}\ ,
\end{align}
where $\hat{c}^{[\alpha,my]}(nm,\ell)$ is the discrete Fourier transform of the Fourier coefficients of the Jacobi forms $\psi^{[g^x,g^y]}(\tau,z)$.  
The  zeroth Fourier-Jacobi coefficient is given in terms of the multiplicative eta product associated with the cycle shape for $g$, i.e., $\rho=1^22^23^26^2$.
\begin{equation}\label{addseeds6}
\begin{split}
\phi_{2,1}^{[g^3,g^2]}(\tau,z) &:= \frac{\vartheta_1(\tau,z)^2}{\eta(\tau)^6} \ \eta_\rho(\tau/3)= \frac{e^{\pi i/6}}4 \phi^{[g,1]}\Big|_{2,1} \gamma_{1/2}\cdot T(\tau,z) \ ,\\
\phi_{2,1}^{[g^2,g^3]}(\tau,z) &:= \frac{\vartheta_1(\tau,z)^2}{\eta(\tau)^6} \ \eta_\rho(\tau/2)= \frac{e^{\pi i/6}}9 \phi^{[g,1]}\Big|_{2,1} \gamma_{1/3}\cdot T(\tau,z)\ ,
\end{split}
\end{equation}
with $\gamma_{1/p} = -ST^{-p}S$.

\subsubsection{The additive lift and the square-root}

The additive lift is given by carrying out the $(\gamma_{1/p}\cdot T)$-transform of the Hecke operator of Cl\'ery-Gritsenko given in Eq. \eqref{CGHecke}. We do not provide explicit formulae for the transformed operator as we did for the S-transform. The square-root continues to make sense as all the Jacobi forms that appear in the product formula have integral coefficients. Further, the zeroth Fourier-Jacobi coefficient has a square-root involving only integral powers of the Dedekind eta function. 

Define the Siegel modular forms $P_1'(\mathbf{Z})$ and $\widetilde{P}_1'(\mathbf{Z})$ via the relations:
\begin{equation}
\Phi_2^{[g^2,g^3]}(\mathbf{Z}) = \Big(P_1'(\mathbf{Z})\Big)^2\quad,\quad
\Phi_2^{[g^3,g^2]}(\mathbf{Z}) = \left(\widetilde{P}_1'(\mathbf{Z})\right)^2\ .
\end{equation}

%
%

\subsubsection[A second additive lift]{A second additive lift\footnote{This section is based on a computation due to Cl\'ery.}}

The additive lift of Cl\'ery as given by Theorem \ref{CleryAL} using the additive seeds given in Eq. \eqref{addseeds6} leads to Siegel modular forms for subgroups of the paramodular group at level $t>1$ which is clearly distinct from the ones we obtain as expansion about other cusps. We will discuss the lift of the square-roots of the additive seeds.
\begin{enumerate}
\item We first consider the lift of the following Jacobi form.
\[
\frac{\vartheta_1(\tau,z)}{\eta(\tau)^3} \sqrt{\eta_\rho(\tau/3)} 
\in J_{1,\tfrac12}(\Gamma_0(2,3),\chi\times v_H) \ .
\] 
This leads to a weight one Siegel modular form for the group $\mathbf{\Gamma}_3^+(6,3,2,1)$.
\item We first consider the lift of the following Jacobi form.
\[
\frac{\vartheta_1(\tau,z)}{\eta(\tau)^3} \sqrt{\eta_\rho(\tau/2)} 
\in J_{1,\tfrac12}(\Gamma_0(3,2)) \ .
\] 
This leads to a weight one Siegel modular form for the paramodular group $\mathbf{\Gamma}_2^+(6,2,3,1)$.
\end{enumerate}
These two Siegel modular forms are clearly distinct from $P_1'(\mathbf{Z})$ and $\widetilde{P}_1'(\mathbf{Z})$. We have not determined the relationship, if any, between these two modular forms.

\subsection{Umbral Moonshine}

Cheng, Duncan and Harvey proposed a generalisation of Mathieu moonshine that associated a variety of objects (finite groups, mock modular forms, Jacobi forms) to each of the 23 Niemeier lattices\cite{Cheng:2012tq,Cheng2014aa}. For instance, they conjectured (proved in \cite{Duncan2015a,Gannon:2012ck}) the existence of an infinite dimensional module for each of the Niemeier lattice such that graded traces correspond to mock modular forms of a particular type. In this section, we will construct Siegel modular forms of the paramodular groups for the subset of Niemeier lattices constructed out of root lattices of $A$-type. In some cases, as we show these also appear in the context of generalised Mathieu moonshine. In all cases that we consider, there is a product formula for the Siegel modular form and for some there is an additive lift as well. We look for cases where the square root makes sense and these turn out to be related to BKM Lie superalgebras as we  will show in the next section.

Let $t|24$. $t=1,2,3,4,6,8,12,24$. Let $m=24/t$. There exists a Siegel modular form of the full paramodular group $\Gamma_t$ that can be obtained in two ways:
\begin{itemize}
\item As a Borcherd's product formula with multiplicative seed, $\varphi_1^{(\ell)}$, which is a Jacobi form of weight zero and index $(\ell-1)$ defined in \cite{Cheng:2012tq}. $\ell$ is referred to as the lambency.
\item As an additive lift with seed $\frac{\theta_1(\tau, z)^2}{\eta(\tau)^6}\times \eta(\tau)^m$ for $t\leq4$. Two other cycle shapes $2^44^4$ and $1^36^3$ correspond to twining by elements $2B$ and $3A$ of the umbral group at lambency three, $2.M_{12}$.
\end{itemize}

\subsubsection{Product Formulae for umbral moonshine}\label{umbralPF}

Theorem \ref{CGproduct} of Cle\'ry and Gritsenko (see appendix \ref{SecCGFormula}) leads to a Borcherds product formula for a Siegel modular form with the input being a Jacobi form. We refer to the Jacobi form as  a multiplicative seed. The Jacobi forms associated with umbral moonshine with lambency $\ell$ are Jacobi  forms of weight zero and index $(\ell-1)$. Column 4 of Table \ref{umbraltable} lists umbral Jacobi forms associated with root lattices with A-type components. 
At lambency 3, we include two more examples that correspond to twining by elements $3A$ and $2B$ of the finite group, $2.M_{12}$, associated with the root lattice $A_2^{12}$. 
The umbral Jacobi forms were constructed using the McKay-Thompson series for these conjugacy classes of  $2.M_{12}$ given in Appendix C of \cite{Cheng:2012tq}. In appendix C, we give explicit formulae for these Jacobi forms.
We obtain Siegel modular forms for the paramodular group at level $(\ell -1)$ i.e.,  $\mathbf{\Gamma}_{\ell-1}$,  (or its sub-group for the two twining examples) whose zeroth Fourier-Jacobi(FJ)  coefficient is listed in Column 3 of Table \ref{umbraltable}. This zeroth FJ coefficient determines the modular properties of Siegel modular form. For instance, the weight of the Siegel modular form is the same as the weight of the zeroth Fourier-Jacobi form.

\subsubsection{Additive Lift for umbral moonshine}

If the Siegel modular form admits an additive life, the zeroth Fourier-Jacobi coefficient given by product formula discussed above in section \ref{umbralPF} must be the additive seed for the additive/arithmetic lift given by Theorem \ref{CleryAL}.  For all the examples that we discuss  with weight $k>0$, this turns out to be true. Let $\phi_{k,1}(\tau,z)$ be the zeroth Fourier-Jacobi coefficient listed in Table \ref{umbraltable}. Let $T_\ell\phi_{k,1}$ be a short form for $\phi|_{k,1} T_-(\ell)(\tau,z)$ where $T_-(\ell)$ is the Hecke operator defined in Theorem \ref{CleryAL}.  Then, 
the equivalence of the additive lift and the multiplicative lift requires
\begin{equation}
\varphi^{\ell}(\tau,z) = \frac{T_\ell\phi_{k,1}(\tau,z)}{\phi_{k,1}(\tau,z)}\ ,
\end{equation}
where $\varphi^{(\ell)}(\tau,z)$ is the weight zero Jacobi form that appears at lambency $\ell$. All examples with $k>0$ also appear as Siegel modular forms for generalized Mathieu moonshine associated with a pair of commuting elements of $M_{24}$. We indicate the corresponding cycle shape as it appears in Mason's list of eta products associated with generalized Mathieu moonshine given in \cite{Mason1990}. The cycle shapes $8^3$ and $24^1$ appear in list of multiplicative eta products given in Dummit et al.\cite{Dummit:1985} but are not associated with $M_{24}$.

\begin{table}[h]
\renewcommand{\arraystretch}{2}
\centering
  \begin{tabular}{|c|c| l|l|}
    \hline
    $\ell$ & $\rho$,  Label  & $~~~$Zeroth FJ Coefficient & Umbral Jacobi Form  \\ \hline
2& $1^{24}$& $\phi_{10,1}=\frac{\theta_1(\tau, z)^2}{\eta(\tau)^6}\times \eta(\tau)^{24}$ & $2\varphi_1^{(2)}= \frac{T_2 \phi_{10,1}}{\phi_{10,1}}$ \\ \hline
& $2^{12}$, $\BZ_2\times \BZ_2A$ & $\phi_{4,1}=\frac{\theta_1(\tau, z)^2}{\eta(\tau)^6}\times \eta(\tau)^{12}$ & $2\varphi_1^{(3)}= \frac{T_3 \phi_{4,1}}{\phi_{4,1}}$ \\ \cline{2-4}
3& $6^32^3$, $\BZ_2\times \BZ_6A$ & $\phi'_{1,1}=\frac{\theta_1(\tau, z)^2}{\eta(\tau)^6}\times \eta(3\tau)^3\eta(\tau)^3$ & $\varphi_1^{(3,3A)}=\frac{T_3 \phi'_{1,1}}{\phi'_{1,1}}$\\ \cline{2-4}
& $2^4 4^4$, $\BZ_2\times \BZ_4E$ & $\phi'_{2,1}=\frac{\theta_1(\tau, z)^2}{\eta(\tau)^6}\times \eta(2\tau)^4 \eta(\tau)^4$ & $2\varphi_1^{(3,2B)}=\frac{T_3 \phi_{2,1}'}{\phi_{2,1}'}$ \\ \hline
4& $3^{8}$, $\BZ_3\times \BZ_3A$ & $\phi_{2,1}=\frac{\theta_1(\tau, z)^2}{\eta(\tau)^6}\times \eta(\tau)^{8}$ & $2\varphi_1^{(4)}= \frac{T_4 \phi_{2,1}}{\phi_{2,1}}$ \\ \hline
5& $4^{6}$, $\BZ_4\times \BZ_4C$ & $\phi_{1,1}=\frac{\theta_1(\tau, z)^2}{\eta(\tau)^6}\times \eta(\tau)^{6}$ & $2\varphi_1^{(5)}= \frac{T_5 \phi_{1,1}}{\phi_{1,1}}$ \\ \hline
7& $6^{4}$ & $\phi_{0,1}=\frac{\theta_1(\tau, z)^2}{\eta(\tau)^6}\times \eta(\tau)^{4}$ & $2\varphi_1^{(7)}= \frac{T_7 \phi_{0,1}}{\phi_{0,1}}-\frac{1}{7}(\phi_{(0,1)})^6 $\\ \hline
9& $8^{3}$& $\phi_{-1/2,1}=\frac{\theta_1(\tau, z)^2}{\eta(\tau)^6}\times \eta(\tau)^{3}$ & $\varphi_1^{(9)}  $\\ \hline
13& $12^{2}$& $\phi_{-1,1}=\frac{\theta_1(\tau, z)^2}{\eta(\tau)^6}\times \eta(\tau)^{2}$ & $2\varphi_1^{(13)}$\\ 
 \hline
 25& $24^{1}$& $\phi_{-3/2,1}=\frac{\theta_1(\tau, z)^2}{\eta(\tau)^6}\times \eta(\tau)$ & $\varphi_1^{(25)}$\\ 
 \hline
    \end{tabular}
    \caption{Siegel modular forms for umbral Moonshine: $\ell$ is the lambency, $\rho$ is a cycle shape that we associate with it and the label, when given, indicates the pair of $M_{24}$ elements that appear in Table 3 of \cite{Mason1990}. The last column gives the weight zero umbral Jacobi form that is also the multiplicative seed. }\label{umbraltable}
\end{table}

\subsection{Taking the square-root}

In Table \ref{umbraltable}, we indicate that the coefficients of the umbral Jacobi form are even integers by writing the umbral Jacobi form as $2\,\varphi$. Further, if the square-root of the zeroth Fourier-Jacobi coefficient also does not involve square-roots of the Dedekind eta function,  there exists a Siegel modular form that is the square-root of the Siegel modular form. This Siegel modular form can be constructed directly using the additive lift with the additive seed given by the square-root of the zeroth FJ coefficient. The multiplicative lift is the same as before with the  seed being one half of the umbral Jacobi form which has integral coefficients. For $(\ell,k)=((2,5),(3,2),(4,1),(5,1/2),(7,0)$, in Table \ref{periodictable}, we denote the modular forms given by the square-root by $\Delta_k(\mathbf{Z})$. 

\subsubsection{Properties of the square-root }

Let $\Phi^{[g,h]}(\mathbf{Z})$ denote the Siegel modular forms  that we have constructed in this section with $h$ having order $N$ and $g$ having order $M$. When  the square-root makes sense, let $\Delta(\mathbf{Z})$ denote the square-root of $\Phi^{[g,h]}(\mathbf{Z})$.
For all $N\in\{1,\ldots,6\}$,  the Siegel modular forms $\Delta(\mathbf{Z})$ transform suitably under the extended S-duality group $W \rtimes \textrm{Dih}(\mathcal{M}_N)$ that was defined in Section \ref{SecLattice}. In these cases, the following properties hold:
\begin{enumerate}
\item For $\gamma^{(N)}$ and $\delta$ which generate $\text{Dih}(\mathcal{M}^{(N)})$, one has
\[
\Delta (\gamma^{(N)}\cdot \mathbf{Z}) =  \Delta(\mathbf{Z})\quad , \quad
\Delta (\delta\cdot \mathbf{Z}) =  \Delta(\mathbf{Z})\ .
\]
$\gamma^{(N)}\cdot \mathbf{Z}:=  (\gamma^{(N)~-1})^{T} \ \mathbf{Z} \ (\gamma^{(N)})^{-1} $ and $\delta \cdot \mathbf{Z}:=  (\delta^{-1})^{T} \ \mathbf{Z} \ \delta^{-1} $.
\item For $w\in W$, 
\[
\Delta (w\cdot \mathbf{Z}) = \det(w) \ \Delta(\mathbf{Z})\ .
\]
For $N=5,6$, $W$ is replaced by $W_+$ -- this is the group generated by elementary reflections from simple roots that belong to $\mathbf{X}_N$.
\item Further,
\[
\Delta(\mathbf{Z}) = \sum_{w\in W} \det(w)\ e^{-w(\varrho^{(N)})} + \cdots 
\]
Again, for $N=5,6$, $W$ is replaced by $W_+$.
\item There is also a product formula
\[
\Delta(\mathbf{Z}) = e^{-\varrho^{(N)}}\ \prod_{\alpha\in L_+} (1- e^{-\alpha})^{\textrm{mult}(\alpha)}\ ,
\]
with $\textrm{mult}(\alpha)=+1$ for all $\alpha\in \mathbf{X}_N$ for $N\leq 4$ and $\alpha\in (\mathbf{X}_N\cup \widetilde{\mathbf{X}}_N)$ for $N=5,6$. The product formula defines the set $L_+$ which we interpret as the set of positive roots. This does not hold for $\Delta_0(\mathbf{Z})$ for which the $\textrm{mult}(\alpha)=-1$  for all roots in $\widetilde{\mathbf{X}}_6$.
\end{enumerate}
\begin{proof}
The properties follow from the following observations. 
\begin{itemize}
\item[(a)] Let $s_0$ refer to the elementary Weyl reflection due to the root $\alpha_0$. This takes $z\rightarrow -z$ under which the additive seed changes sign.  It follows from the additive lift, that the Siegel modular form given by the additive lift also changes sign. A similar conclusion follows from the product formula of Cl\'ery-Gritsenko. Next, the combination $\delta \cdot s_0$ is the element $[1,0,0]$ of the Heisenberg sub-group of $SL(2,\BZ)$ defined in Eqn. \eqref{sl2embedapp}. Under this, the additive seed changes sign. It follows that the Siegel modular form is invariant under $\delta$.
\item[(b)] For the CHL $\BZ_N$ orbifolds, the element $\gamma^{(N)}$ belongs to the level $N$ sub-group of $Sp(2,\BZ)$  (defined in Eq. \eqref{SGroup}) under which the Siegel modular form is invariant. For the examples of umbral moonshine where the Siegel modular forms are invariant under the paramodular group of level $t=N$, again $\gamma^{(N)}$ belongs to the group defined in Appendix \ref{modularstuff}
\item[(c)] Since the action of the dihedral group on $w_0$ generates all elementary reflections generated by real simple roots that appear in $\mathbf{X}_N$, property 2 follows. It is important to emphasise that the Siegel modular form is \textit{not} invariant under $\sigma^{(N)}$ which is defined for $N=5,6$.
\item[(d)] Property 3 follows from property 2 when combined with the presence of the term $e^{-\varrho^{(N)}}$.
\item [(e)] In all the examples that we consider, the Cl\'ery-Gritsenko multiplicative lift holds and hence property 4 holds provided we establish the multiplicities of a subset of the roots.  We do this on a case by case basis.
\end{itemize}
\end{proof}

\begin{table}
 \newcommand\T{\rule{0pt}{3.2ex}}
 \centering
\begin{tabular}{c||c|c|c|c|c|c ||c}\hline
\backslashbox{$N$}{$M$} & $1$ & $2$ & $3$ & $4$ & $5$ & $6$ & Cartan matrix \\ \hline \hline
$1$ &\cellcolor{gray!30} $\stackrel{1^{24}}{\Delta_5}$ & $\stackrel{1^82^8}{\nabla_3}$ & $\stackrel{1^63^6}{\nabla_2}$ & $\stackrel{1^42^24^4}{\nabla_{3/2}}$\T &$\stackrel{1^{4}5^4}{\nabla_1}$ & \cellcolor{blue!30} $\stackrel{1^22^23^36^2}{P_1}$ & $A^{(1)}$ \\[3pt] \hline
$2$ & $\stackrel{1^82^8}{\widetilde{\nabla}_3}$ &\cellcolor{gray!30} $\stackrel{2^{12}}{\Delta_2}$ & \cellcolor{blue!30} $\stackrel{1^22^23^36^2}{P'_1}$ & \cellcolor{gray!30} $\stackrel{2^{4}4^4}{Q_1}$\T & \xcancel{\phantom{\T$\nabla_{3/2}$}} & \xcancel{\phantom{\T$\nabla_{3/2}$}} & $A^{(2)}$ \\[3pt] \hline
$3$ & $\stackrel{1^63^6}{\widetilde{\nabla}_2}$ &\cellcolor{blue!30} $\stackrel{1^22^23^36^2}{\widetilde{P}'_1}$&\cellcolor{gray!30} $\stackrel{3^{8}}{\Delta_1}$\T & \xcancel{\phantom{\T$\nabla_{3/2}$}} &\xcancel{\phantom{\T$\nabla_{3/2}$}}&\xcancel{\phantom{\T$\nabla_{3/2}$}}& $A^{(3)}$\\[3pt] \hline
$4$ & ~$\stackrel{1^42^24^4}{\widetilde{\nabla}_{3/2}}$ & $\stackrel{2^{4}4^4}{\widetilde{Q}_1}$&\xcancel{\phantom{\T$\nabla_{3/2}$}} & \cellcolor{gray!30} $\stackrel{4^{6}~}{\Delta_{1/2}}$\T & \xcancel{\phantom{\T$\nabla_{3/2}$}} & \xcancel{\phantom{\T$\nabla_{3/2}$}} & $A^{(4)}$ \\[3pt] \hline
$5$ & ~$\stackrel{1^45^4}{\widetilde{\nabla}_{1}}$ &\xcancel{\phantom{\T$\nabla_{3/2}$}} &\xcancel{\phantom{\T$\nabla_{3/2}$}} & \xcancel{\phantom{\T$\nabla_{3/2}$}} & \xcancel{\phantom{\T$\nabla_{3/2}$}} & \xcancel{\phantom{\T$\nabla_{3/2}$}} & $A^{(5)}$ \\[3pt] \hline
$6$ &\cellcolor{blue!30}  ~$\stackrel{1^22^23^26^2}{\widetilde{P}_{1}}$ & \xcancel{\phantom{\T$\nabla_{3/2}$}}  &\xcancel{\phantom{\T$\nabla_{3/2}$}} & \xcancel{\phantom{\T$\nabla_{3/2}$}}  & \xcancel{\phantom{\T$\nabla_{3/2}$}} &\cellcolor{gray!30} $\stackrel{6^{4}~}{\Delta_{0}}$& $A^{(6)}$ \\[3pt] \hline
\end{tabular}
\caption{The new periodic table of  BKM Lie superalgebras. The modular forms $P_1'$, $\widetilde{P}_1$ and $\widetilde{P}_1'$ (shaded in blue) can be thought of as expansions of $P_1$ about inequivalent cusps. All entries except for $\Delta_0$ are associated with pairs of commuting elements of $M_{24}$ with $N$ being the order of orbifolding element and $M$ being the order the twining element. The entries in grey are related to umbral mooshine. }
\label{periodictable}
\end{table}

\section{BKM Lie algebras}

In this section, we construct BKM Lie superalgebras whose lattice of real roots are those considered in section \ref{SecLattice} and the Siegel modular forms $\Delta(\mathbf{Z})$ (of section \ref{SecSiegel}) are their Weyl-Kac-Borcherds denominator formulae.

\subsection{Lie algebras from Cartan matrices}

Let $A^{(N)}$ be the Cartan matrices defined in Eq. \ref{CartanMatrix}. For fixed $N$, Let $\mathfrak{g}(A^{(N)})$ represent the following Kac-Moody (KM) Lie algebra in the Chevalley basis. 
The Lie algebra in the Chevalley basis is
\[
 [h_{\alpha_i}, h_{\alpha_j}]=0\ ,\  [h_{\alpha_i},e_{\alpha_j}] = a_{ji} \ e_{\alpha_i}\ ,\   [e_{\alpha_i},e_{-\alpha_i}] = h_{\alpha_i}\ ,
\]
where $A^{(N)}=(a_{ij})$ is the Cartan matrix of the Lie algebra.
The elements of $\mathcal{R}_+$ are generated by multiple commutators of the simple roots subject to the Serre relations:
\[
 \overbrace{[e_{\alpha_i},[e_{\alpha_i},\cdots,[e_{\alpha_i}}^{(1-a_{ij}) \text{ times}},e_{\alpha_j}]\cdots ]] =0\  \text{ for all } i\neq j\ .
\]
For $N>1$, the Cartan matrices are degenerate and one needs to extend them by adding additional roots\cite[Chap. 1]{Kac1985}. This is similar to what is done in the case of the affine $sl(2)$ algebra. For example, the addition of a single root to $A^{(2)}$ would lead to the following non-degenerate Cartan matrix. This corresponds to adding a row and column to $A^{(2)}$.
\begin{equation}
C=\begin{pmatrix}
~~2 & -2 & -6 & -2 & 0 \\
 -2 & ~~2 & -2 & -6 & 0 \\
 -6 & -2 & ~~2 & -2 & 0 \\
 -2 & -6 & -2 & ~~2 & 1 \\
 ~~0 & ~~0 & ~~0 & ~~1 & 0 \\
 \end{pmatrix}
\end{equation}
As is well known, determining the multiplicities of roots for a give KM Lie algebra is not easy. For the affine case, the problem was solved by Macdonald by relating the Weyl character and denominator formulae to Jacobi forms\cite{Macdonald:1972}.

\subsection{The Weyl Denominator Formula}

The Weyl vector $\varrho$ has the property that $(\varrho,\alpha_i) =-\frac12(\alpha_i,\alpha_i)$ for all real simple roots.
It is easy to see that $(s_i(\varrho) -\varrho )=  \alpha_i \in L_+$. More generally, for any $w\in W$, one has $(w(\varrho) -\varrho )\in L_+$.The Weyl-Kac denominator formula is given by 
\[
\sum_{w\in W} \det(w) \  e^{-w(\varrho)}  = e^{-\varrho}\ \prod_{\alpha\in L_+} (1-e^{-\alpha})^{\text{mult}(\alpha)}\ ,
\]
where $L_+$ is augmented by the addition of imaginary roots for affine KM algebras. 
The LHS knows about the real simple roots as they generate the Weyl group. 
The RHS provides details of the space of all roots. However, in general, it is hard to determine the multiplicities of roots. For affine KM algebras, the answer is known by connecting the denominator formula to modular forms\cite{Macdonald:1972,Macdonald:1981}.

 Borcherds addressed this multiplicity problem by adding  imaginary simple roots to KM algebras. Imaginary roots have norm $\leq 0$.
The diagonal elements in the (extended) Cartan matrix now have non-positive entries. The denominator formula gets modified leading to the Weyl-Kac-Borcherds denominator formula
\[
 \Delta=\sum_{w\in W}\text{det}(w)\ w \Big(\sum_{a\cup 0} \epsilon(a) 
 e^{-\varrho-a}\Big)  = e^{-\varrho}\ \prod_{\alpha\in L_+} (1-e^{-\alpha})^{\text{mult}(\alpha)}\ ,
\]
where $a$ is the sum of imaginary simple roots and $\epsilon(a)=(-1)^n$ if $a$ is the sum of $n$ pairwise independent orthogonal roots and zero otherwise.
Borcherds adds imaginary simple roots such that the above sum/product becomes a suitable modular form, $\Delta$ that is referred to as the automorphic correction by Gritsenko and Nikulin. 
Such modular forms admit product formulae (``Borcherd products") leading to an explicit determination of root multiplicities. Gritsenko and Nikulin extended Borcherds idea to the case of superalgebras.

\subsection{Automorphic corrections and a no-go theorem}

In \cite[Sec. 1.4]{Gritsenko:2002}, Gritsenko and Nikulin develop the theory of 
Lorentzian Kac-Moody Lie superalgebras, a class of BKM Lie superalgebras. This is based on their seminal work\cite{GritsenkoNikulinI,GritsenkoNikulinII}. They provide the modifications necessary to include fermionic imaginary simple roots to Borcherds' denominator formula.  The data that they associate with Lorentzian Kac-Moody Lie superalgebras are: (i) A hyperbolic lattice $\mathcal{L}$,  (ii) a reflection group, $W\subset O(\mathcal{L})$, (iii) A set of simple roots $\mathbf{X}$, with a Weyl vector $\varrho$,  that bound a fundamental chamber and (iv) a holomorphic automorphic form on a symmetric domain that provides the automorphic correction to the Lie algebra constructed from the Cartan matrix associated with data (i)-(iii).
 In particular, the ones that appear in  section \ref{SecLattice} with Cartan matrix $A^{(N)}$ for $N=1,2,3$ appear in their classification. These considerations extend to the $N=4$ case, where the Weyl vector is of parabolic type.  Let $\varrho^{(N)}$, $\mathcal{M}^{(N)}$ and $\mathcal{L}^{(N)}$ denote respectively,  the Weyl vector, the hyperbolic polygon and lattice defined in section \ref{SecLattice}. The Weyl denominator formula for Lorentzian Kac-Moody Lie superalgebras takes the form\cite{Gritsenko:2002}
\begin{multline}\label{GNDenominatorFormula}
\Delta(\mathbf{Z}) = e^{-\pi i (\varrho^{(N)},\mathbf{Z}')}\prod_{\alpha \in L_+} \Big(1- e^{-\pi i (\alpha,\mathbf{Z}')}\Big)^{\textrm{mult}(\alpha)}\\
= \sum_{w\in W} \det(w)\, \Big((e^{-\pi i (w(\varrho^{(N)}),\mathbf{Z}')}- \sum_{a\in \mathcal{L}^{(N)} \cap \mathbb{R}_{++}\mathcal{M}^{(N)}} m(a)\,e^{-\pi i(w(\varrho^{(N)}+a),\mathbf{Z}')}\Big) 
\end{multline}
where $\mathbf{Z}'=\left(\begin{smallmatrix} \tau' & -z \\ -z & \tau\end{smallmatrix}\right)= \det(\mathbf{Z})\, \mathbf{Z}^{-1}$ and $W$ is the Weyl group generated by the real simple roots that correspond to the edges of the hyperbolic polygon $\mathcal{M}^{(N)}$.  The Siegel modular form $\Delta(\mathbf{Z})$ determines the multiplicity of positive roots $\textrm{mult}(\alpha)$ as well as the correction due to imaginary simple roots.
In paper I, we showed the existence of several inequivalent automorphic corrections to $\mathfrak{g}(A^{(N)})$ for $N\leq 4$. Some of the modular forms that appeared as the denominator formulae were constructed by Cl\'ery and Gritsenko\cite{Gritsenko:2008}. Others appeared in the papers \cite{Govindarajan:2008vi,Govindarajan:2009qt}. These results were summarised in a periodic table of BKM Lie superalgebras in Paper 1. This table has been updated with several entries as we will discuss next.

The case when the lattices have generalised Weyl vector of hyperbolic type are interesting. There are \textit{no} known examples of BKM Lie superalgebras associated with these lattices. In fact, there is a no-go theorem due to Gritsenko and Nikulin that implies that there are no Lorentzian Kac-Moody Lie superalgebras that satisfy the Weyl denominator formula as given in Eq. \eqref{GNDenominatorFormula}. Needless to say, it is anticipated that there might be new class of BKM Lie superalgebras that circumvent the no-go theorem. We provide evidence in the following sub-section that there exist Macdonald-type identities that are associated with the Cartan matrices $A^{(5)}$ and $A^{(6)}$. We interpret them as Weyl denominator identities for the new class of BKM Lie superalgebras. These appear as rows 5 and 6 of Table \ref{periodictable}. Based on these two examples, we expect to see modifications in the sum side of Eq. \eqref{GNDenominatorFormula}. First, we observe that the modular form transforms covariantly under the Weyl group generated by the simple real roots $\alpha$ whose inner product with Weyl vector is such that $\langle \varrho^{(N)},\alpha\rangle =-1$. Second, even though the other set of real simple roots which satisfy $\langle \varrho^{(N)},\alpha\rangle =+1$ appear on the product side with multiplicity $+1$, we expect that they should provide additional terms that are similar in spirit to how Borcherds modified the sum side when imaginary simple roots appeared. We do not have a precise characterisation of this modification. Thus, we expect to find a new version of Eq. \eqref{GNDenominatorFormula} that works when one has a rank three lattice with Weyl vector of hyperbolic type.

\subsubsection{BKM Lie superalgebras for $A^{(1)}$}

It was shown by us in \cite{Govindarajan:2018ted} that there exist inequivalent automorphic corrections to $\mathfrak{g}(A^{(1)})$ (associated with all conjugacy classes of two distinct $L_2(11)$ sub-groups of $M_{12}$ that we denote by $L_2(11)_A$ and $L_2(11)_B$. In this paper, we focus only on the $L_2(11)_B$ cases and these appear in row 1 of Table \ref{periodictable}. There is also one cycle shape, $1^2 2^1 4^2$ that does not belong to $L_2(11)_B$. The only new entry (as compared to paper 1) corresponds to the modular form $P_1$ in the table.

\subsubsection{BKM Lie superalgebras for CHL $\BZ_N$ orbifolds}

These make up the entries for column 1 of Table \ref{periodictable}. For $N\leq 4$, these have already appeared in paper 1. The new results in this paper are (i) the use of moonshine to obtain  product formulae for Siegel modular forms and (ii) direct formulae for their additive lift.

For $N=5,6$, we have explicitly verified that the additive lift and the multiplicative lift for the modular forms, $\tilde{\nabla}_1$ and $\tilde{P}_1$, agree to fairly high order. It was incorrectly argued in \cite{Krishna:2010gc} that the sum side did not agree with the product formula for $N=5$ orbifold.  This was based on taking the square-root of the generating function of $\frac14$-BPS states as given by \cite{Jatkar:2005bh} and looking for some terms. However, our explicit construction of the additive lift has proven the match of the two constructions. The modularity of the additive lift follows from the modularity of the additive lift of Cl\'ery and Gritsenko. However, the modularity of the product formula that follows from moonshine is not obvious. One needs show that this formula is equivalent the analog of the S-transform of Borcherds product formula of Cl\'ery-Gristenko. An all-orders proof of our claim that goes beyond our checks needs us to prove modularity for the product formula. We have implicitly assumed that this is true\footnote{We thank the anonymous referee for raising this issue.}.  These are the Macdonald type identities for the new kind of BKM Lie superalgebras. 

\subsubsection{BKM Lie superalgebras for Umbral moonshine}

The diagonal entries in Table \ref{periodictable} correspond to Siegel modular forms that arise as multiplicative lifts with seeds given by some of the Jacobi forms that appear in umbral moonshine. The diagonal entry, $\Delta_0$ is a Siegel modular form that provides the automorphic correction to $\mathfrak{g}(A^{(6)})$. The entry corresponding to $Q_1$ also corresponds to twining by an order 2 element in the conjugacy class $2B$ of the umbral group $2.M_{12}$ at lambency 3. 

\subsection{Studying the Macdonald identities for $A^{(5)}$ and  $A^{(6)}$ }

The square-roots of the modular forms that are the generating function of $\BZ_N$ CHL orbifolds for $N=5,6$ provide the first candidates that one can study in order to come up with a modification to the Gritsenko-Nikulin theory of Lorentzian Kac-Moody superalgebras. Once such a modification is developed, one could, in principle, come up with many more examples that potentially arise from Nikulin's examples of rank three hyperbolic lattices with Weyl vector of hyperbolic type.  

 For all roots $\alpha \in \mathbf{X}_N$, one has $(\varrho^{(N)},\alpha)=-1$ while for roots $\tilde{\alpha}\in \tilde{\mathbf{X}}_N$, one has $(\varrho^{(N)},\tilde{\alpha})=+1$. Due to this difference, one has
$[s_\alpha(\varrho^{(N)})-\varrho^{(N)}]=\alpha \in \mathbf{X}_N$ and $[s_{\tilde\alpha}(\varrho^{(N)})-\varrho^{(N)}]=-\tilde{\alpha}\in -\mathbf{X}_N$.
The modular forms transform as expected under only the group generated by all simple roots that appear in $\mathbf{X}_N$. Let us denote this group by $W_+$. One has
\[
\Delta (w \cdot \mathbf{Z}) = \det(w)\ \Delta(\mathbf{Z}),\quad \forall w\in W_+\ .
\]

\subsubsection{Parsing the product formula for $\BZ_N$ CHL orbifolds}


We first focus on the following four simple real roots  for $N=5$
\[
\alpha_0 =\begin{pmatrix} 0 & -1 \\ -1 & 0 \end{pmatrix} \quad ,\quad
\beta_{-1} = \begin{pmatrix} 0 & 1 \\ 1 & 10 \end{pmatrix}\quad,\quad
\tilde{\alpha}_0 =\begin{pmatrix} 4 & 9 \\ 9 & 20 \end{pmatrix} \quad ,\quad
\tilde{\beta}_{-1} = \begin{pmatrix} 6 & 11 \\ 11 &20 \end{pmatrix}\quad.
\]
Similarly, for $N=6$, we focus on the four simple roots
\begin{equation}\label{basicsixroots}
\alpha_0 =\begin{pmatrix} 0 & -1 \\ -1 & 0 \end{pmatrix} \quad ,\quad
\beta_{-1} = \begin{pmatrix} 0 & 1 \\ 1 & 12 \end{pmatrix}\quad,\quad
\tilde{\alpha}_0 =\begin{pmatrix} 2 & 5 \\ 5 & 12 \end{pmatrix} \quad ,\quad
\tilde{\beta}_{-1} = \begin{pmatrix} 4 & 7 \\ 7 &12  \end{pmatrix}\quad.
\end{equation}

Under Weyl reflections due to the four simple roots on each other, one can generate the following twelve real roots (with $c_5=20$ for $N=5$ and $c_6=12$ for $N=6$)
\begin{multline*}
\alpha_0+2\beta_{-1}, 
2\alpha_0+\beta_{-1},\ \alpha_0+(-2+c_N)\, \tilde{\alpha}_0\ , (-2+c_N)\, \alpha_0+\tilde{\alpha}_0,\ \alpha_0+(2+c_N) \tilde{\beta}_{-1},\ (2+c_N)\alpha_0+\tilde{\beta}_{-1}, \\ \beta_{-1}+(2+c_N) \tilde\alpha_{0},  (2+c_N) \beta_{-1} + \tilde{\alpha}_0, \beta_{-1}+2 \tilde\beta_{-1}, \ 2 \beta_{-1} + \tilde{\beta}_{-1}, \ \tilde{\alpha}_0 + 2 \tilde\beta_{-1}\ \text{and }2\tilde\alpha_0 + \tilde\beta_{-1}\ .
\end{multline*}
We have checked  that all these roots appear on the product side with multiplicity $+1$. For the CHL $\BZ_6$ orbifold, the terms in bold below determine the multiplicity of the simple roots $\alpha_0$, $\tilde{\alpha}_0$ and $\tilde{\beta}_{-1}$.
\begin{equation}\label{CHLSixMultiplicity}
\begin{split}
\widehat{\psi}^{[g,1]}=\left(\mathbf{\tfrac{1}{r}} + r \right) + \cdots
+ 
   \left(\tfrac{1}{r^5}+\tfrac{132}{r^4}-\tfrac{1951}{r^3}+\tfrac{9736}{r^2}-\tfrac{23592}{r}+31348+\cdots +\mathbf{r^5}\right) q^6+\cdots \\
+\left(\tfrac{1}{r^7}+\tfrac{660}{r^6}-\tfrac{23592}{r^5}+\tfrac{256724}{r^4}-\tfrac{1344230}{r^3}+\tfrac{4047528}{r^2}-\tfrac{7633817}{r}+9393452+\cdots +\mathbf{r^7}\right) q^{12}\ ,
\end{split}
\end{equation}
where $\widehat{\psi}^{[g,1]}(\tau,z):= \frac1{12} \sum_{a=0}^5 \psi^{[g,g^a]}(\tau,z)$ is the relevant combination as follows from Eq. \eqref{moonshineproduct}.
Hence, all roots that arise from the action of Dih$(\mathcal{M}^{(N)})$, i.e., in particular those in $\mathbf{X}_N$ and $\tilde{\mathbf{X}}_N$, appear with the same multiplicity. 
 
 \subsubsection{The sum side for $\BZ_N$ CHL orbifolds}
 
The formula for the additive lift enables us to study the sum side of the formula. Let $\alpha[n,\ell,m]$ denote the matrix $\left(\begin{smallmatrix} 2n & \ell \\ \ell & 2m\end{smallmatrix}\right)$ and make the formal identification $e^{-\alpha[n,\ell,m]}$ with $q^nr^\ell s^m$.
We make the following observations. 
\begin{enumerate}
\item All simple roots  $\alpha\in\mathbf{X}_N$ appear as $s_\alpha (e^{-\varrho^{(N)}}) = e^{{-\varrho^{(N)}}} e^{-\alpha}$ with coefficient $-1$. 
\item There is no term corresponding $e^{\varrho^{(N)}}$. This is consistent with $\sigma^{(N)}$ not being a symmetry of the Siegel modular form.
\item Further, there are \textbf{no} terms of the form $e^{-\varrho^{(N)}} e^{-\tilde\alpha}$ for any $\tilde{\alpha}\in \tilde{\mathbf{X}}_N$. This suggests that there are amazing cancellations occuring on expanding the product side of the formula.
\item  The $\tilde{\alpha}$ roots appear as $s_{\tilde\alpha} (e^{\varrho^{(N)}}) = e^{-\varrho^{(N)}} e^{2\varrho^{(N)}-\tilde\alpha}$ with a coefficient of $+1$.  
\item The additive seed generates some of the terms involving imaginary simple roots. For $N=5$, let $m(a)$ be defined as follows.
\[
\prod_{n=1}^\infty (1-q^n)^{-1} (1-q^{\frac n5})^2 =1+ \sum_{a=1}^\infty  m(a)\  q^{\frac a5}=1 - 2\, q^{1/5}-q^{2/5}+2q^{3/5} +\cdots\ .
\]
 Then, for all $a\geq 1$, $m(a)$ denotes the multiplicity of the imaginary simple root (given by $\alpha[\frac a5,0,0]$) as it appears in the sum side of the denominator formula Eq. \eqref{GNDenominatorFormula}.. Further, all other imaginary simple roots are given by the action of the Weyl group on this root.

 \item Under the action of symmetries one has:
\[
\cdots \stackrel{\gamma^{(5)}}{\longrightarrow}   q^{16/5} r^8 s^5\stackrel{\gamma^{(5)}}{\longrightarrow} q^{1/5} \stackrel{\gamma^{(5)}}{\longrightarrow} q^{1/5} r^2 s^5 \stackrel{\gamma^{(5)}}{\longrightarrow} \cdots 
\]
All these roots appear with multiplicity $-2$ on the sum side. The generator $\sigma$  acts as follows on a imaginary root
\[
q^{1/5} \stackrel{\sigma^{(5)}}{\longleftrightarrow} q^{4/5} r^2 s^5 
\]
and the root on the right  appears on the sum side with multiplicity $+4$ and not $-2$.  This is consistent with the fact that $\sigma$ is not a symmetry of the modular form.
\item Similar considerations apply for the $N=6$ orbifold but we do not present the details.
 \end{enumerate}


\noindent These are consistent with the following terms on the sum side:
\[
\Sigma = \sum_{w \in \mathcal{W}_+} \det(w)\ w\left( e^{-\varrho^{(N)}}  -\sum_{\eta\in I} m(\eta)\  e^{-\varrho^{(N)} -\eta} + \sum_{\tilde\alpha\in \tilde{\mathbf{X}}_N} e^{\varrho^{(N)}- \tilde\alpha} +\cdots \right)
\]
where sum over $I$ represents the correction terms that appear from imaginary simple roots. The ellipsis indicate that we do not, as yet, have a complete characterization of the sum side.

\subsubsection{The case of $\Delta_0(\mathbf{Z})$}

This is the square-root of the Siegel modular form for umbral moonshine at lambency $7$. We only have a product formula with multiplicative seed given by the weight zero Jacobi form $\varphi_1^{(7)}$ defined in Appendix \ref{umbraljacobi}.
The first few terms in its expansion are:
\[
\varphi_1^{(7)}=\left(\mathbf{\tfrac{1}{r}}+r\right) + \left(-\tfrac{1}{r^5}+\tfrac{1}{r}+r-\mathbf{r^5}\right)\,q 
+   \left(-\tfrac{1}{r^7}-\tfrac{1}{r^5}+\tfrac{2}{r}+2 r-r^5-\mathbf{r}^7\right)\,q^2+\cdots
\]
 On the product side, we find that the roots $\alpha_0$ and $\beta_{-1}$ appearing with multiplicity $+1$ while the simple roots $\tilde{\alpha}_0$ and $\tilde{\beta}_{-1}$ appear with multiplicity $-1$. This is in contrast to what we observe for the $\BZ_6$ CHL orbifold (see Eq. \ref{CHLSixMultiplicity}). These terms are indicated in bold in the above equation. Similarly, $(\tilde{\alpha_0}+2 
 \tilde{\beta}_{-1})$ and $(2 \tilde{\alpha_0}+ 
 \tilde{\beta}_{-1})$ also appears with multiplicity $-1$. Since the Siegel modular form transforms properly under $W_+$, it follows that all terms that lie in the $W_+$ orbit also have multiplicity $-1$.   This is different from what we saw for the case of the $\BZ_6$ CHL orbifold. This suggests that simple roots that appear in $\tilde{\mathbf{X}}_6$ must be treated as fermionic simple roots. Since, we do not have an additive lift, we do not have an independent formula for the sum side. That does not preclude the existence of a BKM Lie superalgebra as was the case in some examples studied in \cite{Govindarajan:2018ted}.

\section{Conclusion}

We have seen how the physics of the walls of marginal stability when combined with generalized moonshine leads to Siegel modular forms that are potential candidates for the Weyl-Kac-Borcherds denominator formulae for a class of Lorentzian Kac-Moody Lie superalgebras in some cases for which the mathematical theory is well developed and a new class of BKM Lie superalgebras for which such a theory doesn't exist. While we do not provide the theory, we have provided some evidence that there might be such a theory associated with hyperbolic root lattices with Weyl vector of hyperbolic type. In \cite{Gritsenko:2002}, Gritsenko and Nikulin express a similar sentiment: ``... \textit{it seems, there is a more general class of Lie algebras (analogous to Lorentzian Kac–Moody algebras which we consider here) such that for this class it is necessary to consider reflective hyperbolic lattices and identities similar to (1.4.14) with a generalized Weyl vector $\rho$ having square with any sign.}" The three examples that we have obtained here might lead to such a theory. This is something we hope to report on in the future\cite{ShabbirVish}.

In \cite{Govindarajan:2018ted}, we have constructed several Lorentzian Kac-Moody Lie superalgebras -- one for every conjugacy class of two inequivalent $L_2(11)$ subgroups of $M_{12}$.  The real simple roots for  all of them have the same Cartan matrix, i.e., $A^{(1)}$. Our extended periodic table of BKM Lie superalgebras has two examples with Cartan matrix $A^{(2)}$ that are related to umbral moonshine at lambency $3$. Are there more? 

We have not considered examples of generalized Mathieu moonshine that reduces to elements of $L_2(11)_A$. We anticipate that the more general product formula given in \cite{Persson:2013xpa} that takes into account phases that arise due to the non-trivial third homology class of $M_{24}$ i.e., $H_3(M_{24},\BZ)=\BZ_{12}$, will be applicable in these cases. \\

\noindent \textbf{Acknowledments:} We thank M. Shabbir and S. Viswanath for useful discussions. We also thank Fabien Cl\'ery for an email correspondence and for providing an English translation of the relevant portions of his thesis. We thank the anonymous referee for numerous suggestions that have improved the presentation of this paper.  SG also thanks the Department of Science and Technology for support through grant EMR/2016/001997 and SS is supported by  a NET-JRF fellowship under UGC fellowship scheme.


\appendix

\section{The paramodular group}\label{modularstuff}

We follow the exposition given in the doctoral thesis of Cl\'ery\cite{Clery2009} that lead to the definition of  paramodular groups with level structures.\\

\noindent\textbf{Sub-groups of} $SL(2,\BZ)$:
\[ 
\Gamma_0(N,M) =\left\{ \left(\begin{smallmatrix}a & b \\ c & d \end{smallmatrix}\right)\in SL(2,\BZ) \Big| c=0\textrm{ mod }N \textrm{ and } b=0 \textrm{ mod }M\right\}
\]
For the group $\Gamma_1(N,M)$, one imposes the additional condition $a=1\textrm{ mod } N$. Note that $\Gamma_0(N)=\Gamma_0(N,1)$ and $\Gamma^0(M)=\Gamma_0(1,M)$. \\



\noindent\textbf{The paramodular group and its subgroups}:\\

For $t$ a positive integer, the classical paramodular group of level $t$,  $\mathbf{\Gamma}_t$, is defined as follows:
\begin{equation}
\mathbf{\Gamma}_t = \left\{
\left(\begin{smallmatrix}   
* & *t & * & * \\[2pt] * & * & * & *t^{-1} \\[2pt] * & *t & * & * \\[2pt] *t & *t & *t & *  
\end{smallmatrix}\right)\in \textrm{Sp}(2,\mathbb{Q})\bigg|\ \textrm{all } * \in \BZ \right\}\ .
\end{equation}
When $t=1$, then $\Gamma_1=\Sp(2,\BZ)\equiv \Gamma^{(2)}$ is the usual  symplectic group.
For positive integers, $N,N_1,L$ and $K$, define the following subset of $\mathbf{\Gamma}_t$
\begin{equation}
\mathbf{\Gamma}_t (N,N_1,L,K)= \left\{
\left(\begin{smallmatrix}   
a_1 & a_2Lt & b_1N_1 & b_2K \\[2pt] a_3L & a_4 & b_3K & b_4N_1t^{-1} \\[2pt] c_1N & c_2NLt & d_1 & d_2L \\[2pt]c_3NLt & c_4Nt & d_3Lt & d_4  
\end{smallmatrix}\right)\bigg|\ a_1,a_4,d_1,d_4\in  (\BZ/N\BZ)^* \right\}\ .
\end{equation}
If $N_1|LKt$, $L|KN$, $K|LN_1$ and $N|Lt$, then $\mathbf{\Gamma}_t (N,N_1,L,K)$ is a subgroup of $\mathbf{\Gamma}_t$. This subgroup is called the paramodular group at level $t$ and level structure $(N,N_1,L,K)$.

Let $N_1|N$. We denote by $\mathbf{\Gamma}^+_t (N,N_1,L,K)=\mathbf{\Gamma}_t (N,N_1,L,K)\cup \mathbf{\Gamma}_t (N,N_1,L,K) V_t$  a normal double extension of $\mathbf{\Gamma}_t(N,N_1,L,K)$ in $\Sp(2,\mathbb{R})$ with
\begin{equation}\label{Vtdef}
V_t = \tfrac1{\sqrt{t}} \left(\begin{smallmatrix} 0 & t & 0 & 0 \\ 1 & 0 & 0& 0 \\
0 & 0 & 0 & 1 \\ 0 & 0 & t & 0 \end{smallmatrix}\right)\ ,
\end{equation}
with $\det(CZ+D)=-1$.
This acts on $\BH_2$ as 
\begin{equation}
(\tau,z,\sigma ) \longrightarrow (t \sigma, z, \tau/t)\ .
\end{equation}
The group $\mathbf{\Gamma}_t ^+(N,N_1,L,K)$ is generated by $V_t$ and its parabolic subgroup
\begin{equation}
\mathbf{\Gamma}^\infty_t (N,N_1,L,K)=\left\{
\left(\begin{smallmatrix}   
a_1 & 0& b_1N_1 & b_2K \\[2pt] a_3L & 1 & b_3K & b_4N_1t^{-1} \\[2pt] c_1N &0 & d_1 & d_2L \\[2pt]0 & 0& 0 & 1  
\end{smallmatrix}\right)\bigg|\ a_1,d_1\in  (\BZ/N\BZ)^* \right\}\ .
\end{equation}
Note that $\mathbf{\Gamma}^+_t (N,1,1,1)$ was called $\Gamma^+_t(N)$ in Paper 1.

The embedding of 
$\left(\begin{smallmatrix} a & b \\ c & d\end{smallmatrix}\right)
\in \Gamma_0(N,N_1)$ in $\mathbf{\Gamma}_t(N,N_1,L,K)$ is given by
\begin{equation}
\label{sl2embed}
\widetilde{\begin{pmatrix} a & b \\ c & d \end{pmatrix}}
\equiv \begin{pmatrix}
   a   &  0 & b & 0   \\
     0 & 1 & 0 & 0 \\
     c &  0 & d & 0 \\
     0 & 0 & 0 & 1  
\end{pmatrix}
\ .
\end{equation}
The above matrix acts on $\BH_2$ as
\begin{equation}
(\tau,z,\sigma) \longrightarrow \left(\frac{a \tau + b}{c\tau+d},\  
\frac{z}{c\tau+d},\  \sigma-\frac{c z^2}{c \tau+d}\right)\ ,
\end{equation}
with $\det(C\mathbf{Z} + D)=(c\tau +d)$. The Heisenberg group, 
$H(\BZ)$, is generated by $Sp(2,\BZ)$ matrices of the form
\begin{equation}
\label{sl2embedapp}
[\lambda, \mu,\kappa]\equiv \begin{pmatrix}
   1   &  0 & 0 & \mu   \\
    \lambda & 1 & \mu & \kappa \\
     0 &  0 & 1 & -\lambda \\
     0 & 0 & 0 & 1  
\end{pmatrix}
\qquad \textrm{with } \lambda, \mu, \kappa \in \BZ
\end{equation}
The above matrix acts on $\BH_2$ as
\begin{equation}
(\tau,z,\sigma) \longrightarrow \left(\tau,\ z+ \lambda \tau  + \mu,\  
\sigma + \lambda^2 \tau + 2 \lambda z + \lambda \mu +\kappa \right)\ ,
\end{equation}
with $\det(C\mathbf{Z} + D)=1$. 

Let $\phi_{k,m}(\tau,z)$ be a Jacobi form of weight $k$ and index $m$. We define the slash operation as follows:
\begin{equation}
\phi\big|_{k,m}\gamma (\tau,z) = \det(\gamma)^{k-1} (c\tau +d)^{-k} e^{ \frac{-2\pi icmz^2}{c\tau+d}} \phi_{k,m}\left(\tfrac{a\tau + b}{c\tau +d}, \tfrac{z}{c\tau+d}\right)
\end{equation}

\subsection{Jacobi Forms}

\subsubsection*{Theta Functions}

The genus-one theta functions are defined by
\begin{equation}
\theta\left[\genfrac{}{}{0pt}{}{a}{b}\right] \left(\tau,z\right)
=\sum_{\ell \in \BZ} 
q^{\frac12 (\ell+\frac{a}2)^2}\ 
r^{(\ell+\frac{a}2)}\ e^{i\pi \ell b}\ ,
\end{equation}
where $a,b\in (0,1)\mod 2$. We define $\vartheta_1 
\left(\tau,z\right)\equiv\theta\left[\genfrac{}{}{0pt}{}{1}{1}\right](\tau,z)$,
$\vartheta_2 
\left(\tau,z\right)\equiv\theta\left[\genfrac{}{}{0pt}{}{1}{0}\right] 
\left(z_1,z\right)$, $\vartheta_3 
\left(\tau,z\right)\equiv\theta\left[\genfrac{}{}{0pt}{}{0}{0}\right] 
\left(\tau,z\right)$ and $\vartheta_4 
\left(\tau,z\right)\equiv\theta\left[\genfrac{}{}{0pt}{}{0}{1}\right] 
\left(\tau,z\right)$.

\subsection{Jacobi forms corresponding to Umbral Moonshine}\label{umbraljacobi}

 The Niemeier root systems containing only $A$-type components and the corresponding weight zero Jacobi forms are as follows\cite{Cheng2014aa}:
 \begin{center}
\begin{tabular}{l |c| c| c|c|c|c|c| c}
\hline Root System & $A_1^{24}$ & $A_2^{12}$ & $A_3^{8}$ & $A_4^{6}$ & $A_6^{4}$ & $A_8^{3}$ & $A_{12}^{2}$ & $A_{24}$\\
\hline
Jacobi Form &$\varphi_1^{(2)}$ & $\varphi_1^{(3)}$ & $\varphi_1^{(4)}$ & $\varphi_1^{(5)}$ & $\varphi_1^{(7)}$ & $\varphi_1^{(9)}$ & $\varphi_1^{(13)}$ & $\varphi_1^{(25)}$\\
\hline
\end{tabular}
\end{center}
where 
\begin{align*}
&\varphi_1^{(2)} = 4( f_2^2 + f_3^2 + f_4^2)=\left(\tfrac{1}{r}+10 +r\right) + \cdots,\\
& \varphi_1^{(3)} = 2(f_2^2 f_3^2 + f_3^2 f_4^2 + f_4^2 f_2^2)=\left(\tfrac{1}{r}+4 +r\right)+\cdots ,\\
& \varphi_1^{(4)} = 4 f_2^2 f_3^2 f_4^2=\left(\tfrac{1}{r}+2 +r\right)+\cdots,
\end{align*}
and $f_i =\vartheta_i(\tau,z)/\vartheta_i(\tau,0)$ for $i\in\{2,3,4\}$.
\begin{align*}
& \varphi_1^{(5)} =\frac14\left(  \varphi_1^{(4)}  \varphi_1^{(2)} - ( \varphi_1^{(3)}  )^2\right)=\left(\tfrac{1}{r}+1 +r\right)+\cdots ,\\
& \varphi_1^{(7)} =\varphi_1^{(3)}  \varphi_1^{(5)} - ( \varphi_1^{(4)}  )^2=\left(\tfrac{1}{r} +r\right)+\cdots ,\\
& \varphi_1^{(9)} =\varphi_1^{(3)}  \varphi_1^{(7)} - ( \varphi_1^{(5)}  )^2=\left(\tfrac{2}{r}-1 +2r\right)+\cdots ,\\
& \varphi_1^{(13)} =\varphi_1^{(5)}  \varphi_1^{(9)} - 2( \varphi_1^{(7)}  )^2=\left(\tfrac{1}{r}-1 +r\right)+\cdots ,\\
& \varphi_1^{(25)} =\frac12\varphi_1^{(5)}  \varphi_1^{(21)} -\varphi_1^{(7)}  \varphi_1^{(19)} +\frac12( \varphi_1^{(13)}  )^2=\left(\tfrac{2}{r}-3 +2r\right)+\cdots ,
\end{align*}
where
\begin{align*}
&\varphi_1^{(21)} = \varphi_1^{(17)} \varphi_1^{(5)} - 2 \varphi_1^{(19)} \varphi_1^{(13)},\\
&\varphi_1^{(19)} = \varphi_1^{(16)} \varphi_1^{(4)} +2 \varphi_1^{(13)} \varphi_1^{(7)} - \varphi_1^{(15)} \varphi_1^{(5)},\\
&\varphi_1^{(17)} = 4\varphi_1^{(13)} \varphi_1^{(5)} - ( \varphi_1^{(9)})^2,\\
&\varphi_1^{(16)} = 2 \varphi_1^{(13)} \varphi_1^{(4)} + \varphi_1^{(10)} \varphi_1^{(7)} - \varphi_1^{(12)}\varphi_1^{(5)},\\
&\varphi_1^{(15)} =  \varphi_1^{(11)} \varphi_1^{(5)} + 6 \varphi_1^{(13)} \varphi_1^{(3)} - \varphi_1^{(12)}\varphi_1^{(4)},\\
&\varphi_1^{(12)} =  \varphi_1^{(8)} \varphi_1^{(5)} + 3\varphi_1^{(10)} \varphi_1^{(3)} - 8 \varphi_1^{(9)}\varphi_1^{(4)},\\
&\varphi_1^{(11)} = 3 \varphi_1^{(7)} \varphi_1^{(5)} + 2\varphi_1^{(9)} \varphi_1^{(3)} - \varphi_1^{(8)}\varphi_1^{(4)},\\
&\varphi_1^{(10)} = \frac12\left( \varphi_1^{(6)} \varphi_1^{(5)} + \varphi_1^{(8)} \varphi_1^{(3)} - 12\varphi_1^{(7)}\varphi_1^{(4)}\right),\\
&\varphi_1^{(8)} =  \varphi_1^{(6)} \varphi_1^{(3)}  - 5\varphi_1^{(4)}\varphi_1^{(5)},\\
&\varphi_1^{(6)} =  \varphi_1^{(2)} \varphi_1^{(5)}  - \varphi_1^{(3)}\varphi_1^{(4)}\ .
\end{align*}
The weight zero Jacobi forms for cycle shape $2^4 4^4$ and $2^3 6^3$ associated with twining with elements in conjugacy classes $2B$ and $3A$, respectively,  of the umbral group at lambency three are  given by
\begin{align*}
\varphi^{(3,2B)}
&=2 \frac{\vartheta_2(\tau,2z)}{\vartheta_2(\tau,0)},\\
\varphi^{(3,3A)}&= \tfrac1{48} \phi_{0,1}(\tau,z)^2 +\tfrac18 E_2^{(3)}(\tau)\phi_{-2,1}(\tau,z)  \phi_{0,1}(\tau,z)-\left( \tfrac{19}{48} E_2^{(3)}(\tau)^2 +\tfrac{25}{12} E_2^{(6)}(\tau)^2\right.\nonumber\\
&\left. -\tfrac1{24} E_2^{(2)}(\tau)E_2^{(3)}(\tau)  +\tfrac5{12} E_2^{(2)}(\tau)E_2^{(6)}(\tau) -\tfrac{65}{24} E_2^{(3)}(\tau)E_2^{(6)}(\tau)  \right) \phi_{-2,1}(\tau,z)^2\ .
\end{align*}

\section{Three constructions of Siegel modular forms}

\subsection{Product formulae from generalized moonshine}\label{Secmoonshineproduct}

In CFT's on a torus on considers the following traces
\begin{equation}\label{GMdef}
\etabox{g}{h} = \Tr_{\mathcal{H}_h}\big (g\ \cdots \big)\ ,
\end{equation}
where $g$ and $h$ are (commuting) symmetries of the CFT and $\mathcal{H}_h$ is a module $h$ twisted sector. We will also denote the same object by $[g,h]$ in a more compact notation. Let $g$ and $h$ denote, for simplicity, discrete symplectic automorphisms of $K3$ that are commute with each other. For instance, consider the elliptic genus in a twisted sector $\mathcal{K}_h$ of an  orbifold of the K3 CFT by the element $h$ twined by the element $g$.  
\begin{equation}\label{twistedtwinedgenera}
\psi_{0,1}^{[g,h]}(\tau,z):= \textrm{Tr}_{\mathcal{K}_h} \Big(g\ q^{L_0-1} r^{J_L}(-1)^{F}\Big)\ ,
\end{equation}
Following \cite{Govindarajan:2011em}(see also \cite{Persson:2013xpa}), define the following twisted Hecke operator
\begin{equation}\label{TwistedHeckeOperator}
 \psi_{0,1}^{[g,h]}~\Big|V_m~(\tau,z) \equiv \frac1m \sum_{ad =m}\sum_{b=0}^{d-1}  \ \psi_{0,1}^{[g^ah^{-b},h^d]}\left(\tfrac{a\tau+b}{d},az\right)\ . 
\end{equation}
The second-quantized elliptic $h$-twisted genus twined by the element $g$ is defined to be
\begin{equation}
\mathcal{E}^{[g,h]}(\mathbf{Z}):=\exp\left[-\sum_{m=1}^\infty s^m \ \psi_{0,1}^{[g,h]}~\Big|V_m~(\tau,z)\right]
\end{equation}
The Siegel modular form is obtained by
\begin{equation}\label{PhiGeneral}
\Phi_k^{[g,h]}(\mathbf{Z}):= s\ \phi_{k,1}^{[g,h]}(\tau,z) \ \mathcal{E}^{[g,h]}(\mathbf{Z})\ ,
\end{equation}
where
\[
\phi_{k,1}^{[g,h]}(\tau,z) = \frac{\theta_1(\tau,z)^2}{\eta(\tau)^6}\times \eta^{[g,h]}(\tau)\ .
\]

Specialising to the case when both when we consider a twisting by $g^y$ and twining by $g^x$ where $g$ is an element of order $N$, we obtain the following product formula starting from Eq. \eqref{TwistedHeckeOperator}.
\begin{align}\label{moonshineproduct}
\Phi_k^{[g^x,\,g^y]}(\mathbf{Z})= s\phi_{k,1}^{[g^x,\,g^y]} \times \prod_{m=1}^{\infty} \prod_{\alpha=0}^{N-1}\prod_{n\in\mathbb{Z}-\frac{\alpha y}{N}}\prod_{\ell\in \mathbb{Z}}  (1- e^{\frac{2\pi i \alpha x}N}q^{n}r^{\ell} s^{m})^{c^{[\alpha,my]}(nmN,\ell)}\ .
\end{align}

\subsection{Borcherds product formula of Cl\'ery-Gritsenko}\label{SecCGFormula}

\begin{theorem}[Cl\'ery-Gritsenko\cite{Gritsenko:2008}] \label{CGproduct} Let $\psi$ be a nearly holomorphic Jacobi form of weight $0$ and index $t$ of $\Gamma_0(N)$. Assume that for all cusps, $e/f \in \mathcal{P}$, of $\Gamma_0(N)$ one has $\frac{h_e}{N_e} c_{f/e}(n,\ell)\in \mathbb{Z}$ if $4nt -\ell^2\leq 0$. Then the product
  \[
  B_\psi(\mathbf{Z}) = q^A r^B s^C \prod_{f/e\in \mathcal{P}} \prod_{\substack{n,\ell,m\in \mathbb{Z}\\ (n,\ell,m)>0}} \Big(1-(q^nr^\ell s^{tm})^{N_e}\Big)^{\frac{h_e}{N_e} c_{f/e}(nm,\ell)}\ ,
  \]
 where $(n,l,m)>0$ implies: if $m>0$, then $n\in \BZ$ and $\ell\in \BZ$; if $m=0$ and $n>0$, then $\ell \in \BZ$; if $m=n=0$, then $\ell <0$ and
  \[
  A=\frac1{24} \sum_{\substack{f/e\in \mathcal{P}\\ \ell \in \mathbb{Z}}} h_e\, c_{f/e}(0,\ell) ,\ 
  B=\frac1{2} \sum_{\substack{f/e\in \mathcal{P}\\ \ell \in \mathbb{Z}_{>0}}} \ell h_e\, c_{f/e}(0,\ell),\ 
   C=\frac1{4} \sum_{\substack{f/e\in \mathcal{P}\\ \ell \in \mathbb{Z}}} \ell^2 h_e\, c_{f/e}(0,\ell)\ ,
  \]
  defines a meromorphic Siegel modular form of weight 
  \[
  k=  \frac1{2} \sum_{\substack{f/e\in \mathcal{P}\\ \ell \in \mathbb{Z}}} \frac{h_e}{N_e} c_{f/e}(0,0)
  \]
  with respect to $\mathbf{\Gamma}_t(N)^+$ possibly with character. The character is determined by the zeroth Fourier-Jacobi coefficient of $B_\psi(\mathbf{Z})$ which is a Jacobi form of weight $k$ and index $C$ of the Jacobi subgroup of $\mathbf{\Gamma}_t(N)^+$.
  \end{theorem}
  \noindent \textbf{Remark:} As discussed by Cl\'ery and Gritsenko, the poles and zeros of $B_\psi$ lie on rational quadratic divisors defined by the Fourier coefficients $ c_{f/e}(n,\ell)$ for $4n-\ell^2\leq0$. The condition $\frac{h_e}{N_e} c_{f/e}(n,\ell)\in \mathbb{Z}$ ensures that one has only poles or zeros at these divisors. 
  

\subsection{Additive/Arithmetic lift of Cl\'ery-Gritsenko}

We state below a theorem of Cl\'ery that is a generalization of the additve lift of Cl\'ery and Gritsenko.

\begin{theorem}[Special case of Theorem 3.5 of Cl\'ery\cite{Clery2009}] \label{CleryAL} Let $\phi\in J_{k,t}(\Gamma_0(N,N_1),\chi\times v_H^{2t})$ and $k\in \BN$, $t\in \BN/2$, $N_1|N$ and $\chi: \Gamma_0(N) \rightarrow \BC^*$ is a character of finite order such that Ker$(\chi)\supset \Gamma_1(Nq_1,q_1)$ and $N_1|q_1$.  Assume that $q_1$ is a divisor of $24$ and $\chi\left(\left(\begin{smallmatrix} 1 & 1\\ 0 & 1 \end{smallmatrix}\right)\right) = e^{2\pi i /q_1}$.
 If $q_1t\in \BN$ or $q_1=1$ and $c(0,0)=0$ where $c(0,0)$ is the constant coefficient in the Fourier expansion of $\phi$ at the cusp at infinity.   Then the function
$$
F_\phi(Z) = \sum_{\substack{m\equiv 1 \textrm{ mod } q_1\\ m>0}} \widetilde{\phi}|_k T_-^{(N)}(m)(Z)
$$
is a modular form of weight $k$ for the group $\mathbf{\Gamma}_{q_1t}^+(N,N_1,L,1)$ with character $\chi_t$ and $L$ is a positive integer such that $N|Lq_1t$ and $L|N$. 
\end{theorem}
 The character $\chi_t,$ is induced by the character $\chi_\mu\times v_H^{2t}$ of the Jacobi group and the relations
$$
\chi_t(V_{q_1t})=(-1)^k,\quad \chi_{t,\mu}([(0,0);\tfrac{\kappa N_1}{q_1t}]) = e^{2\pi i \tfrac{ \kappa N_1}{q_1}}\quad (\kappa \in \BZ)\ .
$$
\section{Simplifying the Hecke operators for the additive lift}\label{SimplifyHecke}

\subsection{$N=5$}
For $N=5$, the additive seed is given by 
$$ \phi_{1,1/2} (\tau,z) = \frac{\theta_1(\tau,z)}{\eta(\tau)^3} \eta(\tau)^2 \eta(5\tau)^2\ .$$
We compute this Jacobi form at different cusps and they are given by 
\begin{align*}
 \phi  \big|_{1,1/2} S\, (\tau,z)& =-\frac15 \frac{\theta_1(\tau,z)^2}{\eta(\tau)^3} \eta(\tau)^2 \eta(\tau/5)^2=:\tilde{\phi}_{1,\tfrac12}(\tau,z)\ ,\\
\phi \big|_{1,1/2} \gamma_{2/5}\,(\tau,z)& =   -\phi_{1,1/2} (\tau,z) \ ,\\
\phi\big|_{1,1/2}\gamma_{4/5}\,(\tau,z)& =   \phi_{1,1/2} (\tau,z) \ ,
\end{align*}
Then we have
\begin{align*}
\phi \big|_{1,1/2} T_-(5)\cdot S (\tau,z)= & \tilde{\phi} \big|_{1,1/2}\left( \begin{smallmatrix} 5 & 0 \\ 0& 1\end{smallmatrix}\right) (\tau,z) + \phi \big|_{1,1/2} \left( \begin{smallmatrix} 1 & 2 \\ 0& 5\end{smallmatrix}\right)(\tau,z)+ \phi \big|_{1,1/2} \left( \begin{smallmatrix} 1 & -2 \\ 0& 5\end{smallmatrix}\right)(\tau,z) \\ 
&\quad- \phi \big|_{1,1/2} \left( \begin{smallmatrix} 1 & 1\\ 0& 5\end{smallmatrix}\right)(\tau,z) - \phi \big|_{1,1/2} \left( \begin{smallmatrix} 1 & -1 \\ 0& 5\end{smallmatrix}\right)(\tau,z)\\
=& \tilde{\phi}_{1,1/2} (5\tau,5z) +\frac15 \sum_{b=1}^4 \phi_{1,1/2} \left(\tfrac{\tau+b}5,z\right)
\end{align*}
For the last four terms in the first line, the phases inside the arguments of $\phi_{1,1/2}(\tau,z)$ and the signs outside combines to give the second term in the second line. Using  the identity \[
\sum_{b=0}^{4}\tilde{\phi}_{1,1/2} \left(\tfrac{\tau+b}5,z\right)  = -\frac15  \phi_{1,1/2} \left(\tfrac{\tau}5,z\right)\ ,
\]
we obtain
\begin{equation}
\boxed{
\phi \big|_{1,1/2} T_-(5)\cdot S (\tau,z)= \tilde{\phi}_{1,1/2} (5\tau,5z) +\sum_{b=0}^{4}\tilde{\phi}_{1,1/2} \left(\tfrac{\tau+b}5,z\right) +\frac15 \sum_{b=0}^4 \phi_{1,1/2} \left(\tfrac{\tau+b}5,z\right)
}
\end{equation}
which is a formula that is easier to implement in software as it corresponds to picking a particular subset of coefficients in the Fourier expansion.

\subsection{$N=6$}
 For $N=6$, the additive seed is given by 
$$ \phi_{1,1/2} (\tau,z) = \frac{\theta_1(\tau,z)^2}{\eta(\tau)^3} \eta(\tau) \eta(2\tau) \eta(3\tau) \eta(6\tau)\ .$$
We compute this Jacobi form at different cusps and they are given by 
\begin{align*}
\phi \big|_{1,1/2}S\,(\tau,z) &= -\frac{1}{6} \frac{\theta_1(\tau,z)^2}{\eta(\tau)^3} \eta(\tau) \eta(\tau/2) \eta(\tau/3) \eta(\tau/6)=: \tilde{\phi}_{1,1/2}(\tau,z)\\
\phi \big|_{1,1/2}\gamma_{2/3}\,(\tau,z) &= \frac{1}{2} \frac{\theta_1(\tau,z)^2}{\eta(\tau)^3} \eta(3\tau) \eta(\tau) \eta(3\tau/2) \eta(\tau/2)=:\frac12{\phi}_{1,1/2}'(\tau,z)\\
\phi \big|_{1,1/2}\gamma_{4/3}\,(\tau,z) &= \frac{1}{2} \frac{\theta_1(\tau,z)^2}{\eta(\tau)^3} \eta(3\tau) \eta(\tau) \eta(3\tau/2) \eta(\tau/2)=:\frac12{\phi}_{1,1/2}'(\tau,z)
\end{align*}
Then we have
\begin{align*}
\phi \big|_{1,1/2}T_-(3).S\,(\tau,z)  = & \tilde{\phi}  \big|_{1,1/2}\left( \begin{smallmatrix} 3 & 0 \\ 0& 1\end{smallmatrix}\right)\,(\tau,z) + \frac12\phi'  \big|_{1,1/2} \left( \begin{smallmatrix} 1 & 1 \\ 0& 3\end{smallmatrix}\right)\,(\tau,z)+ \frac12\phi' \big|_{1,1/2} \left( \begin{smallmatrix} 1 & -1 \\ 0& 3\end{smallmatrix}\right)\,(\tau,z) \\ 
&=  \tilde{\phi}_{1,1/2} (3\tau,3z)+\frac13\sum_{b=1,2}\frac12 \phi_{1,1/2}'\left(\tfrac{\tau+b}3,z\right)\\
&=  \tilde{\phi}_{1,1/2} (3\tau,3z)+\sum_{b=0}^2 \tilde{\phi}_{1,1/2} \left(\tfrac{\tau+b}3,z\right) +\frac16\sum_{b=0}^2 \phi_{1,1/2}' \left(\tfrac{\tau+b}3,z\right)\ .
\end{align*}
Using the identity $\sum_{b=0}^{2}\tilde{\phi}_{1,1/2} \left(\tfrac{\tau+b}3,z\right)  = -\frac16  \phi_{1,1/2}' \left(\tfrac{\tau}3,z\right)$, we can simplify the formula to 
\begin{equation}
\boxed{
\phi \big|_{1,1/2}T_-(3).S\,(\tau,z)  =  \tilde{\phi}_{1,1/2} (3\tau,3z)+\sum_{b=0}^2 \tilde{\phi}_{1,1/2} \left(\tfrac{\tau+b}3,z\right) +\frac16\sum_{b=0}^2 \phi_{1,1/2}' \left(\tfrac{\tau+b}3,z\right)
}\ .
\end{equation}

\bibliography{master}
\end{document}